\documentstyle[11pt,epsf,psfig,rotating]{elsart}
\def\etal{{\rm et al.~}}

\def\sub{{substructures}}
\def\apim{{$\Pi^{(m)}$}}
\def\api2{{$\Pi^{(2)}$}}
\def\api3{{$\Pi^{(3)}$}}
\def\api4{{$\Pi^{(4)}$}}
\def\h80{{$h_{80}^{-1}$}}
\def\L{{$\Lambda$CDM}}
\def\M{{CHDM}}
\def\gr{\kern 2pt\hbox{}^\circ{\kern -2pt K}} 

\renewcommand{\thefigure}{\arabic{figure}}
\renewcommand{\thetable}{\arabic{table}}
\def\ltsima{$\; \buildrel < \over \sim \;$}
\def\simlt{\lower.5ex\hbox{\ltsima}}
\def\gtsima{$\; \buildrel > \over \sim \;$}
\def\simgt{\lower.5ex\hbox{\gtsima}}

\begin{document}
\begin{frontmatter}
\title{Global cluster morphology and its evolution: \\
X--ray data $vs$ CDM, \L~ and \M~ models}

\author[trieste]{{\thanksref {valda}} Riccardo Valdarnini}
\thanks[valda]{E--mail: valda@sissa.it}
\author[milano,infn]{{\thanksref {simo}} Simona Ghizzardi}
\thanks[simo]{E--mail:  ghizzardi@mi.infn.it}
\author[milano,infn]{{\thanksref {silvio}} Silvio Bonometto}
\thanks[silvio]{E--mail: bonometto@mi.infn.it}

\address[trieste]{SISSA -- International School for Advanced Studies,
    Via Beirut 2/4, Trieste, Italy}
\address[milano]{Department of Physics G.~Occhialini of the University
of Milano--Bicocca}
\address[infn]{INFN -- Sezione di Milano, Via Celoria 16, 20133 Milano, Italy}

\begin{abstract}
The global structure of galaxy clusters and its evolution
are tested within a large set of TREESPH simulations aimed to
allow a systematic statistical comparison with available
X--ray data. Structure tests are based on the so--called
{\it power ratios} introduced by Buote $\&$ Tsai. The cosmological
models considered are flat CDM, \L~($\Omega_\Lambda = 0.7$)
and \M~($\Omega_h = 0.2$, 1 massive $\nu$). All models
are normalized so to provide a fair number density of clusters.
For each model we perform a P3M cosmological simulation in
a large box, where we select the volumes where the most massive
40 clusters form. Then we go back to the initial redshift and
run a hydrodynamical TREESPH simulation for each of them.
In this way we can perform a statistical comparison of the global 
morphology of clusters, expected in each cosmological model,
with ROSAT data, using the Student t--test, the F--test and
the Kolmogorov--Smirnov test.
The last test and its generalization to 2--dimensional distributions
are also used to compare the joint distributions
of 2 or 3 power ratios. We find that using DM
distribution, instead of gas, as was done in some of previous
analyses, leads to systematically biased results, as
the baryon distribution is substantially less structured
than DM distribution. We also find that the cosmological models considered
have different behaviours in respect to these tests:
\L~has the worst performance. CDM and the \M~mixture considered
here have similar scores. Although the general trend of
power ratio distributions is already fit by these models,
a further improvement is expected either
from a different DM mix or a non--flat CDM model.
\end{abstract}
\end{frontmatter}

\section{Introduction}
\label{sec:introduction}
 
Clusters of galaxies are the largest bound structures in the Universe. 
Within the hierarchical clustering scenario they were assembled through
the merging and fragmentation of smaller size objects, which had
collapsed first, within regions $ \simeq 10$--$30\, $Mpc  wide.
Accordingly, their masses are estimated in the range
$ \simeq 10^{14}- 10^{15}  M_{\odot}$.
Their gravitational growth is dominated by collisionless dark matter (DM)
and the gas distribution follows that of DM.

The outputs of such growth depend on a number of cosmological parameters.
First of all, if the global matter density parameter $\Omega_m < 1$, the 
growth of primeval fluctuations is drastically slowed down before the 
present epoch. In particular, if $\Lambda = 0$ (vacuum energy density
$\rho_\Lambda = 0$), a transition from decelerated to steady 
cosmological expansion occurs at a redshift slightly later 
than $ \Omega_m^{-1}$ and this redshift ($z_{tr}$) 
approximately sets an end to fluctuation growth.
If $\Lambda \neq 0$, instead, a freeze of fluctuation growing takes place
after vacuum energy density $\rho_\Lambda$ begins to dominate on matter 
energy density $\rho_m$. As $\rho_m \propto (1+z)^3$, $z_{tr} \sim 
\Omega_m^{-1/3}$. Also in this case, fluctuations not close enough 
to turn around by $z_{tr}$ are doomed to quite a slow growing at later 
epochs. Furthermore, even apart of such late stops to fluctuation growth,
a determinant role is played by the shape of the {\it transferred} 
fluctuation spectrum, which depends on other cosmological parameters,
besides of $\Lambda$ and $\Omega_m$. Henceforth cosmological parameters 
shed their influence over the number and space distribution of clusters
(see, $e.g.$, White, Efstathiou $\&$ Frenk 1993; Eke, Cole $\&$ Frenk 1996),
over their evolution (see, $e.g.$, Jing $\&$ Fang 1994; Bahcall, Fan $\&$ 
Cen 1997; Henry 1997), as well as over their morphology. 

However, in the last case, hydrodynamics 
governs gas distribution, gravitation itself is intrinsically non--linear and,
a priori, it is hard to state which observed features shall be related to 
unavoidable kinematical effects and which else are due to initial conditions 
set up by primeval spectra or, however, to cosmological parameters. 
It is therefore clear why attempts to trace back cosmological parameters 
from data were often concentrated on super--cluster scales.

Finding a suitable parameter set to quantify the global cluster morphology 
is a non trivial task. In this work we shall do so using the power ratios
\apim, introduced by Buote $\&$ Tsai (1995, hereafter BT95), which
are essentially related to a multipole expansion accounting for
the angle dependence of cluster surface brightness. 
We shall see that this is an effective and synthetic
way to discriminate cluster features, as the \apim\
do depend on the cosmological model and discriminate among different
cosmologies. It is quite likely,
however, that they do not exhaust the characteristics of clusters
which depend on the model. Henceforth, both before and after
their introduction, many attempts were performed to provide further
useful statistical tools.

Early works in this field tried to find a relation between the radial density
profile $\rho_{cl}(r)$ and the value of $\Omega_o$. Cen (1994) and  Mohr \etal
(1995) obtained a steeper density profile for open models than for critical
density. However, later analyses (Jing \etal 1995; Crone \etal 1997; Eke,
Cole $\&$ Frenk 1996; Thomas \etal 1998) showed that the radial density 
profile, when scaled to the cluster virial radius, is substantially 
independent from $\Omega_o$. 

There can be little doubts that these are significant quantities, but their
quantitative analysis can be performed only using really wide observational and
model samples, in order to span the whole functional space in an appropriate
way. 

A more promising way to constrain the cosmological model arises from the study
of substructures in the inner mass distribution. This approach has
observational support both from internal galaxy distribution (Geller $\&$ Beers
1982; Dressler $\&$ Shectman 1988; West $\&$ Bothun 1990; Bird 1995) and from
X--ray image brightness (Jones $\&$ Forman 1992; B\"ohringer 1993; Mohr,
Fabricant $\&$ Geller 1993). 

Observed substructures were first compared with those expected within a
spherical growth approximation (Richstone, Loeb $\&$ Turner 1992), then with
those expected within a Press--Schechter approach, to follow the merger history
of subclumps (Kauffman $\&$ White 1993; Lacey $\&$ Cole 1993). Cosmological
simulations, however, are the most natural method to compare substructure
evolution in different models. Early simulations (Evrard \etal 1993; Mohr 
\etal 1995) showed that the fraction of substructure in galaxy clusters is 
negligible in low--density CDM models, thus favoring a cosmology with 
$\Omega_o = 1$. Jing \etal (1995), instead, found a large fraction of galaxy
clusters with substructures in \L\ models with $\Omega_m \simeq  0.3$. 

An attempt to find a more synthetic parameterization of the degree of
inhomogeneity was made by Crone, Evrard \& Richstone (1996), who used the
displacement of the centre of mass as a function of the overdensity level to
quantify the amount of cluster substructures in different cosmological models.
Henceforth, they conclude that the center of mass analysis is a good test to
discriminate among different models, while previous tests suggested by 
Fitchett \& Webster (1987) and Dressler $\&$ Shectman (1988) do not perform so
well. 

Other statistical approaches, trying to quantity substructures, are the moment
of the distributions (Dutta 1995) and the hierarchical clustering method (Serna
$\&$ Gerbal 1996, Gurzadyan $\&$ Mazure 1998). They all confirm that the level
of substructure of galaxy clusters is quite sensitive to the underlying
cosmological models, although their capacity to discriminate among them is
often rather ambiguous. In our opinion, as already outlined, the most promising
method is based on the so--called {\it power ratios} and amounts to a multipole
expansion of the two--dimensional potential generating the observed
surface brightness.

Further detail on their definition and significance are given in the next
section, where we shall also discuss the observational material on which
their analysis can be based. As a matter of fact, complete maps of optical
surface brightness or, let alone, projected mass density (obtained, $e.g.$,
using gravitational lensing) are unavailable. Henceforth Buote \& Tsai (1996,
hereafter BT96) used maps of X--ray surface brightness, which
trace the squared baryon density distribution.

Previous work in this field, besides of using available observational material,
made also use of outputs of simulations already performed for different aims.
They amounted to 6 hydro simulation for CDM model, and to a set of N--body
simulations for various cosmological models. 
The effectiveness of the approach, however, calls for better
observational material and simulations. In this paper we try to fulfill such
requirements for what concerns simulations, running three sets of 40
cluster models, for different cosmological models, with a TREESPH code.
We compare their outputs with observations using, first of all,
the statistical tests used in previous work.
However, some statistical tests will also be suitably improved, obtaining
a higher discriminatory power.

It is also important to stress that we compared our results, based
on baryon distributions, with those obtainable from the same clusters,
if DM distributions are used. We find that, systematically, structures
in gas are less pronounced than in DM. Replacing N--body
simulations with TREESPH simulations is therefore a substantial
step forward and does lead to a different score for the cosmological
models considered.

In our opinion, however, a final word on model rating, based on \apim\ 
analysis, cannot yet be said. This is not only due to the limits of available
observational material, whose biases we try to reproduce in our model analysis.
Rather, some of improved statistical tests considered show no agreement between
data and any of the cosmological model considered. In spite of that, we can
safely state that, from this kind of analysis, \L~ models come out disfavored;
standard CDM and the CHDM mix considered here (which, unlike CDM, fits also
COBE quadrupole data and has a fair spectral shape parameter $\Gamma$) are
certainly closer to observations. However, in our opinion a general fit of data
can be expected for some different DM mix different and/or for open CDM 
models.

The plan of the paper will be given now. In sec. \ref{sec:power} 
the definition
of power ratios will be reviewed and we will also outline the correlation
between power ratios and cluster evolution. Some details about the different
cosmological models considered in our analysis will be reported in sec.
\ref{sec:models}. In sec. \ref{sec:simulations} we give suitable 
details on the
P3M N--body code for the cosmological simulations and the TREESPH for the
hydrodynamical simulations. The statistical tools used for the comparison
between clusters worked out from each cosmological model and observational data
will be summarized in sec. \ref{sec:results}, where the results will 
also be
discussed. Finally, in sec. \ref{sec:conclusions} we present our 
conclusions. 

\section{Power ratios: definition and evolution; observational sample}
\label{sec:power}

We assume that the X--ray intensity received by $ROSAT$ PSPC
(Pfeffermann \etal 1987) arises from 
thermal bremsstrahlung and, therefore, is $I_X = \Lambda \rho_b^2 ({\bf r})$.
In principle, the coefficient $\Lambda$ depends
both on the temperature $T_X$ and on the geometry
of the source. However, the number of photons collected by $ROSAT$ PSPC
does not depend on $T_X$, provided that $T_X \simgt 1\, $keV
(see, $e.g.$, NRA 91--OSSA--3, Appendix F: $ROSAT$ mission description). 
Furthermore, $I_X$ shall be used in quantities whose {\it ratios}
will be compared with data and, therefore, the absolute value of
$\Lambda$ does not appear in the final expressions.

Henceforth, the values of \apim\ from a model cluster are obtained as follows
(here we shall report a procedure fully consistent with BT95, BT96, Tsai
$\&$ Buote 1996, hereafter TB, and Buote $\&$ Xu 1997, hereafter BX):
(i) $\rho_b^2 ({\bf r})$ [which is $\propto  I_X ({\bf r})$] is projected 
on a plane $\pi_r $, chosen at random, to yield the $X$--ray surface 
brightness $\Sigma (x,y)$ (the direction $\hat n$, orthogonal to $\pi_r$, 
defines the {\it line of sight}; $x$ and $y$ are cartesian coordinates on 
$\pi_r$).
(ii) The mass center $O$ of $\Sigma (x,y)$ ({\it centroid}) is used as origin
of plane polar coordinates on $\pi_r$, to obtain $\Sigma (R,\varphi)$. 
(iii) By solving the Poisson equation:
\begin{equation}
\label{Poisson}
\nabla^2 \Phi = \Sigma (R,\varphi) 
\end{equation}
we obtain the pseudo--potential $\Phi(R,\varphi)$
(constant factors, in front of $\Sigma$, would again simplify).
(iv) We expand $\Phi$ in plane harmonics; the coefficients
of such expansion will be used to build the power ratios {\apim}$(R)$
as follows:
\begin{equation}
\label{prdef}
\Pi^{(m)} (R) = \log_{10} (P_m/P_0)
\end{equation}
where
\begin{equation}
\label{mpdef}
P_m (R) = {1 \over 2 m^2} (\alpha_m^2 + \beta_m^2) ~,
P_0 = [ \alpha_0 \ln(R/{\rm kpc}) ]^2
\end{equation}
and
\begin{equation}
\label{int}
\alpha_m = \int_0^1 { ds\, s^{m+1} \int_0^{2\pi} { d\varphi [ \Sigma (sR,
\varphi) R^2] \cos(m\varphi) }}
\end{equation}
\begin{equation}
\beta_m = \int_0^1 { ds\, s^{m+1} \int_0^{2\pi} { d\varphi [ \Sigma
(sR,\varphi) R^2] \sin(m\varphi)}}
\end{equation}
Here one can directly see that constant factors -- unspecified hereabove --
cancel out when ratio of $P_m$'s are taken. 

Owing to the definition of the centroid $O$,
$P_1$ vanishes. Henceforth, also $\Pi^{(1)} \equiv 0$. Furthermore,
we shall restrict our analysis to $m 
\leq 4$, to account for \sub~ on scales not much below $R$ itself,
and this leaves us with only 3 significant \apim ($m=2,3,4$).

Because of its evolution, a cluster moves along a curve of the 3--dimensional
space spanned by such \apim's; this curve is called {\sl evolutionary track}.
We shall also consider its 2--dimensional projections. Quite in general, a
cluster starts from a configuration away from the origin, corresponding to a
large amount of internal structure and evolves towards isotropization and
homogeneization. This motion, however, does not occur with a steady trend:
sudden bursts of structure appear, when matter lumps approach the cluster
potential well, and eventually fade as lumps are absorbed by it. Gradually,
however, the evolutionary track approaches the origin, as can be easily
appreciated by averaging over the contributions of several clusters. 

Actual data, of course, do not follow the motion of a given cluster along the
evolutionary track. Different clusters, however, lie at different redshifts and
can be used to describe a succession of evolutionary stages. Power ratios for
model clusters are to be set in the 3--dimensional space spanned by \apim's,
taking each model cluster at redshifts distributed as for data clusters. 

Our data set is the same used by BT96. Among X--ray cluster images, taken with
$ROSAT$ PSPC, and such that the PSPC central ring contains a cluster portion
whose radius exceeds 400$\, h^{-1}$kpc, we shall use those contained in the
HEASARC--legacy database and belonging to the Ebeling (1993) or Edge \etal
(1990) samples. Out of the 59 objects selected in this way, one can estimate
\apim\ for 44 of them at $R = 0.8\, h^{-1}$Mpc and for 27 of them at $R = 1.2\,
h^{-1}$Mpc. PSPC data give the X--ray surface brightness $\Sigma_X
(R,\varphi)$, which is to be used in the same way as the $\Sigma$'s obtained
from models, to work out \apim\ for $R=0.4,\, 0.8\, ,1.2\, h^{-1}$Mpc. 
According to TB, the resulting sample is partially incomplete, but, clusters 
were not selected for reasons related to their morphology and the missing 
clusters are expected to have a distribution of power ratios similar to the 
observed one.
They also tested this point using results of the Imaging Proportional Counter
of the Einstein satellite. We shall later report on the lack of any similar
correlation for model clusters. 

In Fig. 
1 we report the redshift distribution of clusters in
our data set, for the three values of $R$ we use. We took this into account in
our comparison with simulations, 
 at variance with previous analyses. As we
shall better detail in the next section, for each cosmological model we
consider, we have outputs for 40 simulated clusters, at a set of redshifts
$z_{in}$ (see below). 
\begin{figure}
\vfill
\centerline{\mbox{\epsfysize=7.0truecm\epsffile{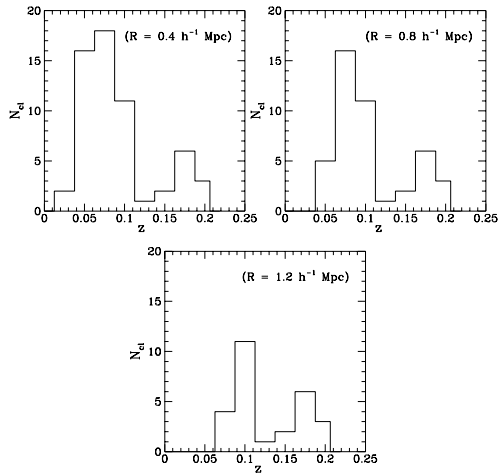}}}
\caption{ Redshift distribution of data clusters for 
$R=0.4,0.8,1.2 h^{-1} {\rm Mpc}$. }
\label{fig:redhis}
\end{figure}
This enables us to select simulation outputs so to reproduce the
$z$--distribution of data using each model cluster only once, $i.e.$ at a
single $z$ value. This can be done in a large number of ways. Among them we
randomly select 50 {\sl cluster sequences}. However, even at this point, we can
still choose the line of sight in each model cluster in each {\sl sequence } in
an arbitrary way. This is fixed at random, once for all, in each cluster of
each sequence. The cluster sequence selection is independently made for each
cosmological model. 

\section{The cosmological models}
\label{sec:models}

In this work we considered three spatially flat cosmological models, that we
shall indicate CDM, \L\ and \M. The Hubble parameter, normalized to 100$\, {\rm
km}\, {\rm s}^{-1} {\rm Mpc}^{-1}$, is $h=0.5$ for CDM and \M, and $h=0.7$ for
\L; for all models the primeval spectral index $n=1$ and the baryon density
parameter is selected to give $\Omega_b h^2 = 0.015$. In \M\ we have 1 massive
neutrino with mass $m_\nu = 4.65\, $eV, yielding a HDM density parameter
$\Omega_h = 0.20$. In \L~ the vacuum contribution to the energy density is
$\Omega_\Lambda = 0.7$. Accordingly, the global density parameter of matter
$\Omega_m$ is 1 for CDM and CHDM, and 0.3 for \L. 

All models were normalized using their transfer function, so to give 32
clusters of mass $M_c > 4.2\, h^{-1} \cdot 10^{14} M_\odot$ in a box of side $L
= 200 \, h^{-1}$Mpc (White \etal 1993, Biviano \etal 1993, Eke \etal 1996,
Girardi \etal 1998). For \M\ and \L\ this gives quadrupole values within 
$2 \sigma$ from COBE 4--year data (Bennet \etal 1996). As is
known, CDM fails this test, but is however an important reference model. For
CDM, also the shape parameter $\Gamma = 7.13 \cdot 10^{-3} \left( \sigma_8 /
\sigma_{25} \right)^{10/3}$ ($\sigma_{8,25}$ are mass variances on $8,25 ~
h^{-1}$Mpc scales) widely exceeds the observational interval. For APM galaxies,
in fact, Peacock and Dodds (1994) found $\Gamma =0.23 \pm 0.04$; for the
Abell/ACO sample, Borgani \etal (1997) found $\Gamma$ in the interval
0.18--0.25; on the contrary CDM predicts $\Gamma \sim 0.4\, $. \M\ and \L,
instead, yield $\Gamma \sim 0.2 $, well inside the observational interval. 

The selection of cosmological models was made in order to have simple models
with significantly different characteristics. Both for \M\ and \L, different
parameter choices are possible. In particular \M\ features are highly dependent
on the DM mix considered. The mix taken here is selected both for its physical
significance and to ease simulations. In fact, \M~ with light neutrinos is non
trivial to simulate, as thermal neutrino velocities rapidly displace hot
particles from their initial positions and the hot spectrum on small scales
becomes dominated by shot noise. With the parameter choice performed here and
in the simulation section, 
this effect is almost absent in this work. 

\section{The simulations}
\label{sec:simulations}

In order to achieve a safe statistical basis for our analysis, we built 40
simulated clusters for each cosmological model considered. Herebelow we report
the details of the simulations, which were a significant part of our work. The
procedure starts from taking a large simulation box, whose side ($200 h^{-1}
{\rm Mpc}$) is fixed in order to provide more than 40 clusters, with mass $ >
10^{14} h^{-1} M_\odot$. In this box we run a (purely gravitational) N--body
P3M simulation for each cosmological model, starting from a redshift $z_{in}$.
Then, using a friend--of--friend (FoF) algorithm, we located the 40 most
massive clusters for each model. For each of them, restarting from $z_{in}$, we
performed a hydrodynamical TREESPH simulation in spheres of radius
$\sim 15$--$25\, h^{-1}{\rm Mpc}$). 

Let us first describe the cosmological N--body simulations. The PM part of the
code makes use of $256^3$ cells. For the PP part of the code the
Plummer--equivalent softening distance is $330 {\rm kpc}$ for all the models.
The particle number is $N_p = 10^6$  for CDM and \M~ models with
$\Omega_m=1$, while $N_p=84^3$ for \L~ with $\Omega_m=0.3$ (see Table 1).

The simulations were run from an initial expansion 
factor
($a_{in}=1$), when $\sigma_8 \simlt 0.1$, to a final expansion factor $a_0$,
when the cumulative cluster number density $n_c(> M_c) = 4 \cdot 10^{-6} h^3
{\rm Mpc}^{-3}$ for $M_c = 4.2 h^{-1} 10^{14} M_\odot$. In CDM and \L~ models
this is expected for $\sigma_8= 0.52\Omega_m^{-0.52+0.13 \Omega_m}$ (White 
\etal 1993; Biviano \etal 1993; Eke \etal 1996) and in CHDM models for 
$\sigma_8 = 0.57$ (Valdarnini, Kahniashvili \& Novosyadlyj 1998). This 
procedure yields $z_{in}=a_0-1$, which must fulfill two competing requests:
(i) In order to apply the Zel'dovich approximation, the linear Lagrangian 
theory must still be a valid approximation; this sets a lower limit to 
$z_{in}$, which depends on the model. (ii) However, the average particle 
separation decreases as the scale factor $a(t)$ and, at $z_{in}$, it risks to 
approach the gravitational softening length which, in TREESPH simulations, 
has a constant physical size.
The values of $a_0$ given in Table 1 have been chosen in order
to fulfill both constraints. Just for the sake of comparison, in their CDM
simulations, Navarro, Frenk \& White (1995, hereafter NFW) have $a_0 = 4.74$. 

\begin{table}
\label{tab:simpar}
\caption{Cosmological simulation parameters}
\begin{center}
\begin{tabular}{cccccc}
models& $\Omega_m$ & h &  $a_0$  & $N_p$ &$\sigma_8$ \\ 
\hline
 CDM & 1 & 0.5  &  4.5 &  $10^6$ & 0.6  \\
 CHDM & 1 & 0.5 &  5.8 & $10^6$ & 0.64 \\
 $\Lambda$CDM & 0.3 & 0.7 & 11  & $84^3$ & 1.1 \\
\hline
\end{tabular}
\end{center}
\vglue 0.2truecm
\par\noindent
{\it Note }: {\small For CHDM, the density parameters are $\Omega_c = 0.8$ 
and $\Omega_h = 0.2$ (1 massive $\nu$).}
\end{table}

At variance from CDM and \L, in CHDM the substance is made of three components.
However, for the P3M simulation, this bears scarce relevance. In fact, the
neutrino Jeans mass at $z_{in}$ is $M_{J\nu}=2 \cdot 10^{13} M_\odot$, just
above particle mass. This allows us to use a global transfer function
${\cal T}(k) = \Omega_c {\cal T}_c(k) + \Omega_h {\cal T}_h(k)$,
where ${\cal T}_{c,h}(k)$ are the transfer functions for the cold and hot 
components, with a suitable growing coefficient $\alpha$ (Klypin \etal 1993). 

Let us also outline that the same random numbers were used to set the initial 
conditions for all three cosmological models. 

\begin{figure}
\label{fig:nmc}
\caption{Cosmological mass functions of galaxy clusters for 
the three considered models. The three curves have been obtained 
at $z = 0$, see Table 1.}
\centerline{\mbox{\epsfysize=7.0truecm\epsffile{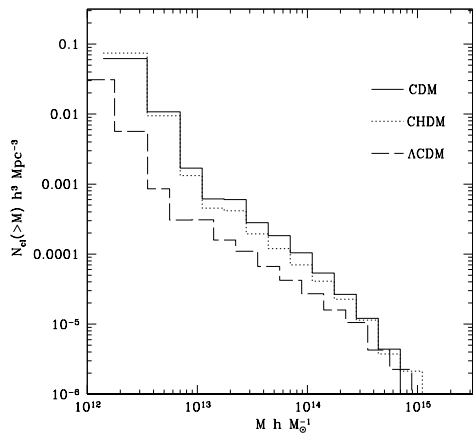}}}
\end{figure}

The FoF algorithm considers as friends the particles at distances smaller than
$0.2\Omega_m^{0.2}$ times the mean particle separation. The linking parameter
is scaled with $\Omega_m$ in order to detect overdensities $\simeq 200
\Omega_m^{-0.6}$. Clusters with mass $ > 4.9\, h^{-1} \cdot 10^{13} M_\odot$
were taken; clusters whose centers--of--mass lie within an Abell radius ($R_A =
1.5 h^{-1} {\rm Mpc}$) are merged together. The three models yielded a spectral
amplitude slightly above linear on the $8 \, h^{-1}$Mpc scale; the values of
$\sigma_8$ which gave the required $n(>M_c)$ are reported in Table
1, together with other parameters for the models 
considered.
The cluster mass distributions are given in Fig. 2.

For the 40 most massive clusters of each cosmological model, a hydrodynamic
simulation was performed in physical coordinates, using a TREESPH code
(Hernquist $\&$ Katz 1989, hereafter HK). In order to do so, we first located
the center of each cluster at $z = 0$ and found all particles within $r_{200}$
(where the cluster density is $\simeq 200\Omega_m^{-0.6}$ times the background
density). These cluster particles were then located at $z_{in}$, in the 
original simulation cube. 

According to a top--hat model we expect that a cube centered on the cluster 
center, and enclosing all its particles at $z_{in}$, has a side 
$L_c \simeq 12 r_{200} \Omega_m^{-0.2}$. However, we found that this
value is often too small to enclose all particles found within $r_{200}$.
In most cases the discrepancy is $\sim 10-30\%$, but, sometimes, it
is much greater than so.
In all cases we set $L_c$ so to enclose all particles  found within $r_{200}$.
$L_c$ values defined in this way range from $15$ to $ 25 h^{-1}$ Mpc. 

A high resolution lattice of $N_L=22^3$ grid points was set in such
cubes. Each grid point corresponds to a mass $m_{grid}=2.7 \cdot 10^{11}
\Omega_m h^2 (L_c/{\rm Mpc})^3 M_{\odot}/N_L$. Such mass is shared in two or
three parts, to obtain the masses $m_i={\Omega_i \over \Omega_m} m_{grid}$
($i={\rm cold,hot,baryon}$) of particles describing the cosmic substance (of
course, $\Sigma_i \Omega_i = \Omega_m$). Different lattices were used for each
substance component, separated by suitable fractions of particle spacings along
each spatial coordinate. For \M\ simulations a small peculiar velocity is given
to the hot particles, drawn at random in magnitude and direction from a
Fermi--Dirac distribution with $v_0=5(1+z_{in})(10 eV/m_{\nu})$km$\, $s$^{-1}$.
For the gas particles we set an initial temperature $T_i=10^4 \gr$. 

The particle  positions given by the high resolution lattices 
were then perturbed, using the same initial conditions of the original 
cosmological simulations, implemented by additional waves to
sample the increased Nyquist frequency. Baryon fluctuations are given by the
same transfer function as CDM. In \M~simulations, neutrinos are perturbed
according to their transfer functions.

However, such cubes are too small to neglect the action of matter
laying outside them. Henceforth, each
cube of side $L_c$ is located inside a greater cube of side $2L_c$,
with parallel sides and equal center, whereinside we set matter yielding
gravitational action, but expected not to take part to the cluster
collapse. Accordingly, gas needs not to be used there and collisionless
particles representing the cold spectrum will account also for baryons.
Also a lesser resolution is sufficient and the greater cube is therefore
filled with a lattice of $N_L$ grid points (their spacing is therefore
double than that for the inner cube).  Accordingly, the mass related to each 
grid point is $8 m_{grid}$. Particles are set only in the part of the greater 
cube ouside the smaller one and their positions and masses are determined 
with the same procedure used for the inner cube.

Here again, however, the particle positions were then perturbed, using 
the same initial conditions of the original cosmological simulations, again
implemented by (a smaller number of) additional waves to sample the 
increased Nyquist frequency. Hence, on the boundaries of such
greater box, we match the initial conditions of the cosmological P3M.

The high resolution TREESPH simulation will then be carried on in physical 
coordinates,   
using all particles which lie inside a sphere of radius $L_c$, and with the
same center as cubes. One must therefore bear in mind that further
gravitational actions, which might have been caused from outside such 
external sphere, are necessarily neglected. This technique is similar to the 
one adopted by Katz \& White (1993) and NFW.

It must be however outlined that, in spite of the above prescriptions,
after running the TREESPH simulation down to $z=0$, we found a few
clusters for which the mass distribution within $r_{200}$ was
contaminated by particles of the external shell. This feature is to be 
avoided, in order to prevent an anomalous two--body heating;
therefore, these particles are identified and a new integration is 
performed with each of these particles split into 8 hot or cold DM 
sub--particles. For the CDM model, the origin of the contamination 
is fairly clear; contaminating particles were those which,
after perturbing their position at $z_{in}$, to account for
the initial spectrum of perturbations, were displaced inside
a radius $L_c/2$ from the cube center.
Unfortunately such rule of thumb is not so reliable for \M~ and \L~ models,
 the difference is presumably related to the different shapes
of power spectra.

The gravitational softening parameter $\varepsilon$ are the same,
independently of the cluster mass, in simulations for a given
cosmological model (apart of 5 simulations in \L; see below). 
They are different, instead, for the different
substance components, according to a law $\varepsilon \propto m_p^{1/3}$ 
($m_p$ is the particle mass of a given substance component), mostly
in order to avoid 2--body heating between different substance components.
This eases other dynamical problems and
ensures a constant central density for all particle species 
(Farouki $\&$ Salpeter 1982).

For gas particles the softenings are $\varepsilon_g=80,\, 100\, ,60\, $kpc 
for CDM,~\M~and most \L~ models, respectively. For the highest
mass 5 \L~ clusters, $\varepsilon_g$ was set up to 80$\, $kpc.
For the CDM component, henceforth, $\varepsilon_c = 200,231,125\, ({\rm or}~
160)\, $kpc for CDM,~\M~and \L~ models, respectively. Accordingly, the
ratio $\varepsilon_g / r_{200} $ never exceeds $\simeq 0.04$.
The actual force resolution of our hydrodynamic simulations compares with
$\varepsilon = 100\, $kpc in NFW, although they do not specify
the substance or $m$ dependence of the softening parameters. 

The hydro/N--body integrations are performed with a tolerance parameter
$\theta=0.7$, without quadrupole corrections.  Viscosity was set as in HK
(Eq.2.22) taking the values  $\alpha = 1$ and $\beta = 2$. We did not consider
the effects of the heating and cooling in the simulations. As a matter of fact,
also previous simulations used in \apim\ analyses were subject to this
restriction, which causes densest regions structures not to be fairly modeled.
However, we do not need to inspect such detailed structures and the restriction
has also an advantage as, to a good approximation, results can be directly
scaled to other $\Omega_b$ values. 

The time integration is done using a standard leap--frog scheme, allowing
each particle $p$ to vary its own time step. In general, it will amount to
$\Delta t_p = \Delta t_0/2^{k_p}$ ($k_p \ge 0$ is the time--bin index of 
the particle $p$). Here $\Delta t_0=8.8 \cdot 10^7 yr$ and
the minimum timestep allowed is $3 \cdot 10^6 yr $ for gas particles and twice 
as much for collisionless ones. 

The value of $k_p$ is selected according to the values taken
by $p$--particle velocity ($v_p$), acceleration ($a_p$) and binding
energy per unit mass ($E_p$), by requiring that
\begin{equation}
a_p v_p \Delta t_p \leq \eta E_p~
\end{equation}
(see Katz 1991). Here $\eta$ is a tolerance parameter
that we set to 0.02$\, $. We ran several integrations with 
the  improved criteria
\begin{equation}
\Delta t_p \leq 0.3 ( \varepsilon_p/a_p)^{1/2}
\end{equation}
\begin{equation}
\Delta t_p \leq 0.3 ( \varepsilon_p/v_p)~,
\end{equation}
but we did not notice appreciable differences in numerical outputs,
presumably thanks to the small value of $\eta$ we used.

In addition to the above criterions the timesteps for gas particles must also
satisfy the Courant stability constraint, so that sound waves cannot change in
a single timestep over the SPH smoothing scale. This constraint reads $ \Delta
t_p \leq \Delta t_C$. Here $\Delta t_C$ has a fairly complex expression, given,
$e.g.$, in Eq.(2.37) of HK. 

In average, each cluster simulation required $\sim 6\, $hours of CPU 
time to be evolved from $z_{in}$ to $z=0$ on a RS6000 computer workstation.
Outputs from TREESPH simulations were preserved at various redshifts.
Among the smaller ones, we used $z_i = $0.15, 0.10, 0.049 and 0
($i=1,..,4$). 

\begin{figure*}
\centerline{\hbox{%
\psfig{figure=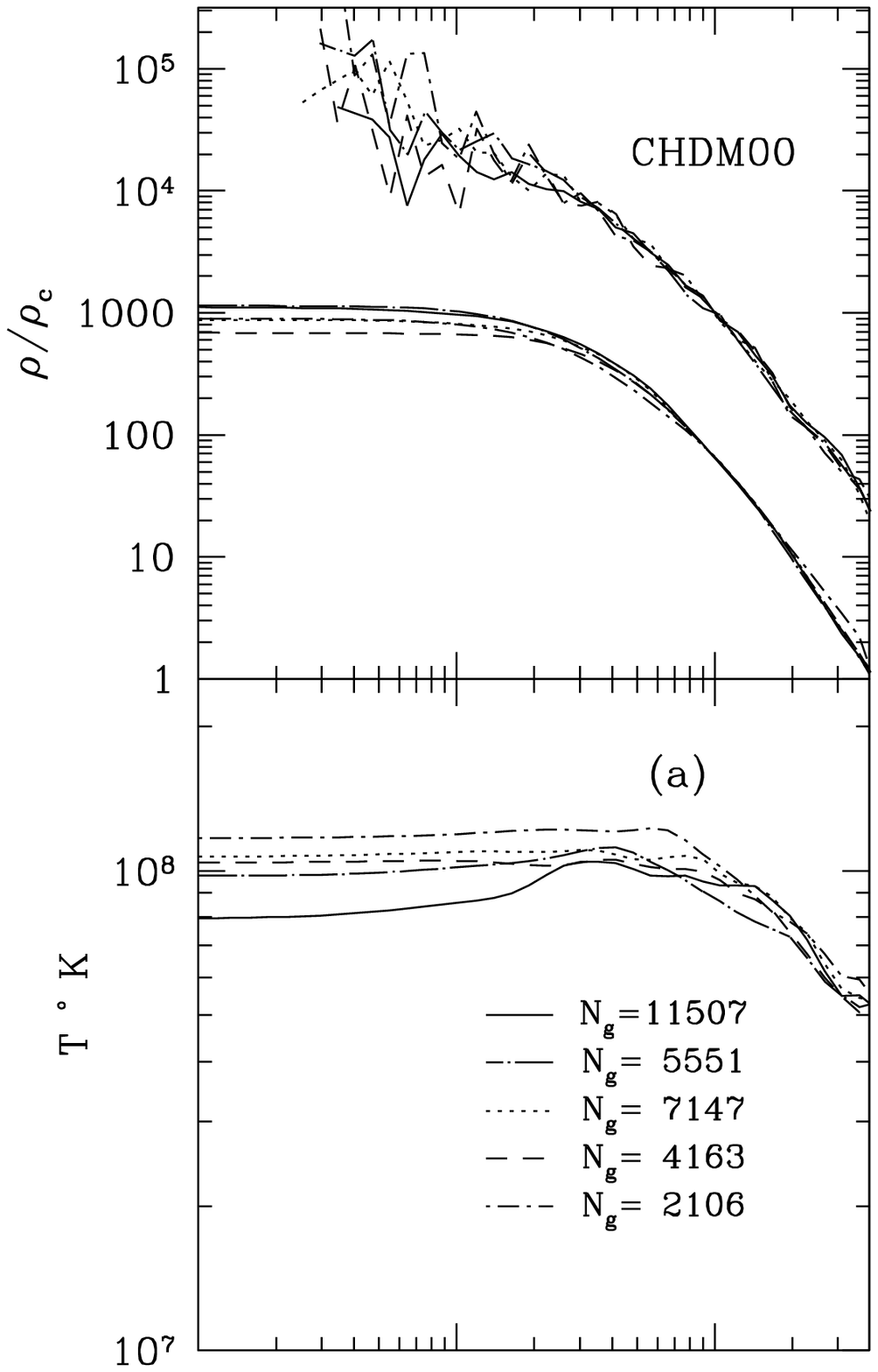,height=11truecm,width=8truecm}%
\psfig{figure=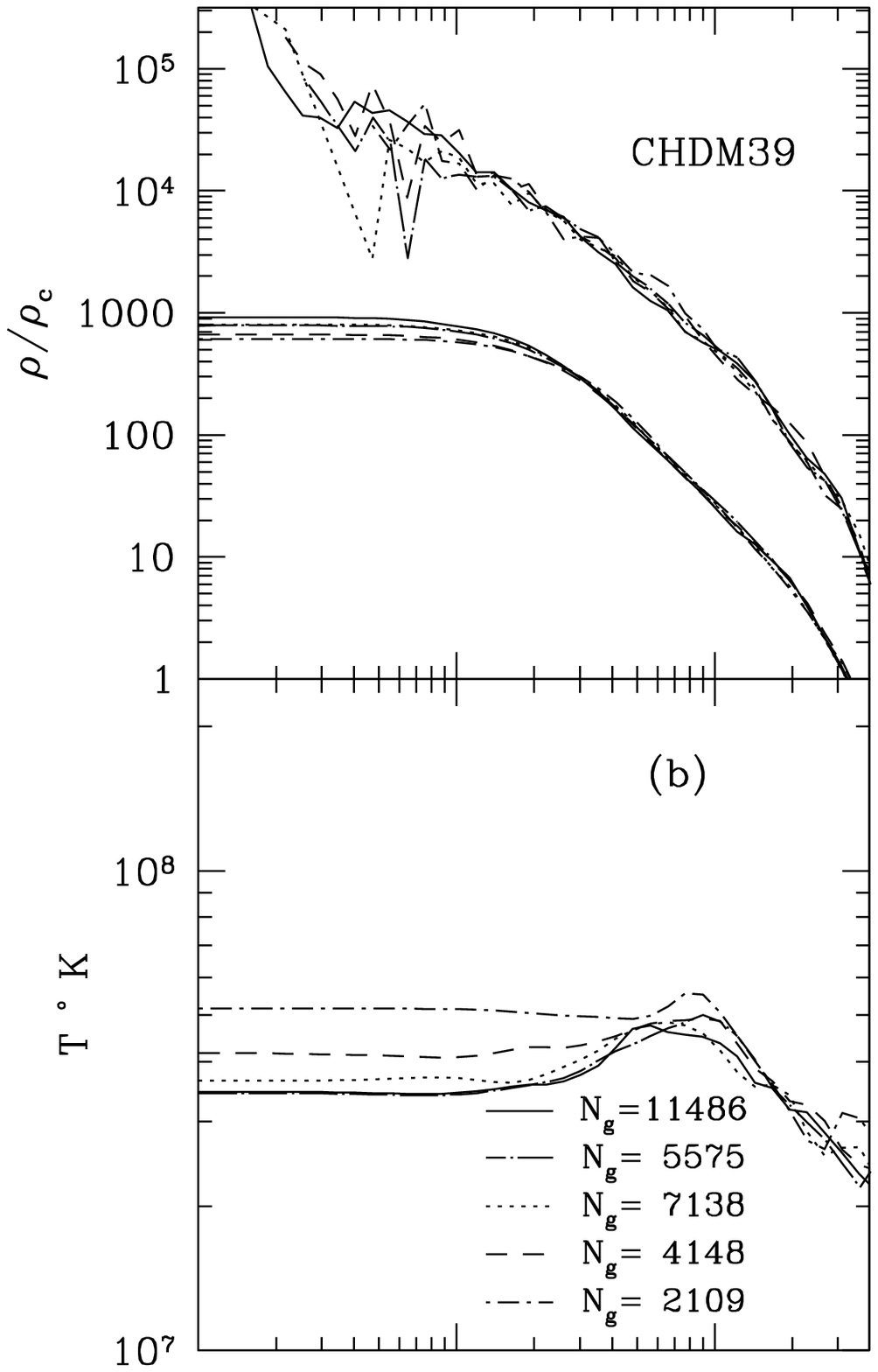,height=11truecm,width=8truecm}%
}}
\centerline{\hbox{%
\psfig{figure=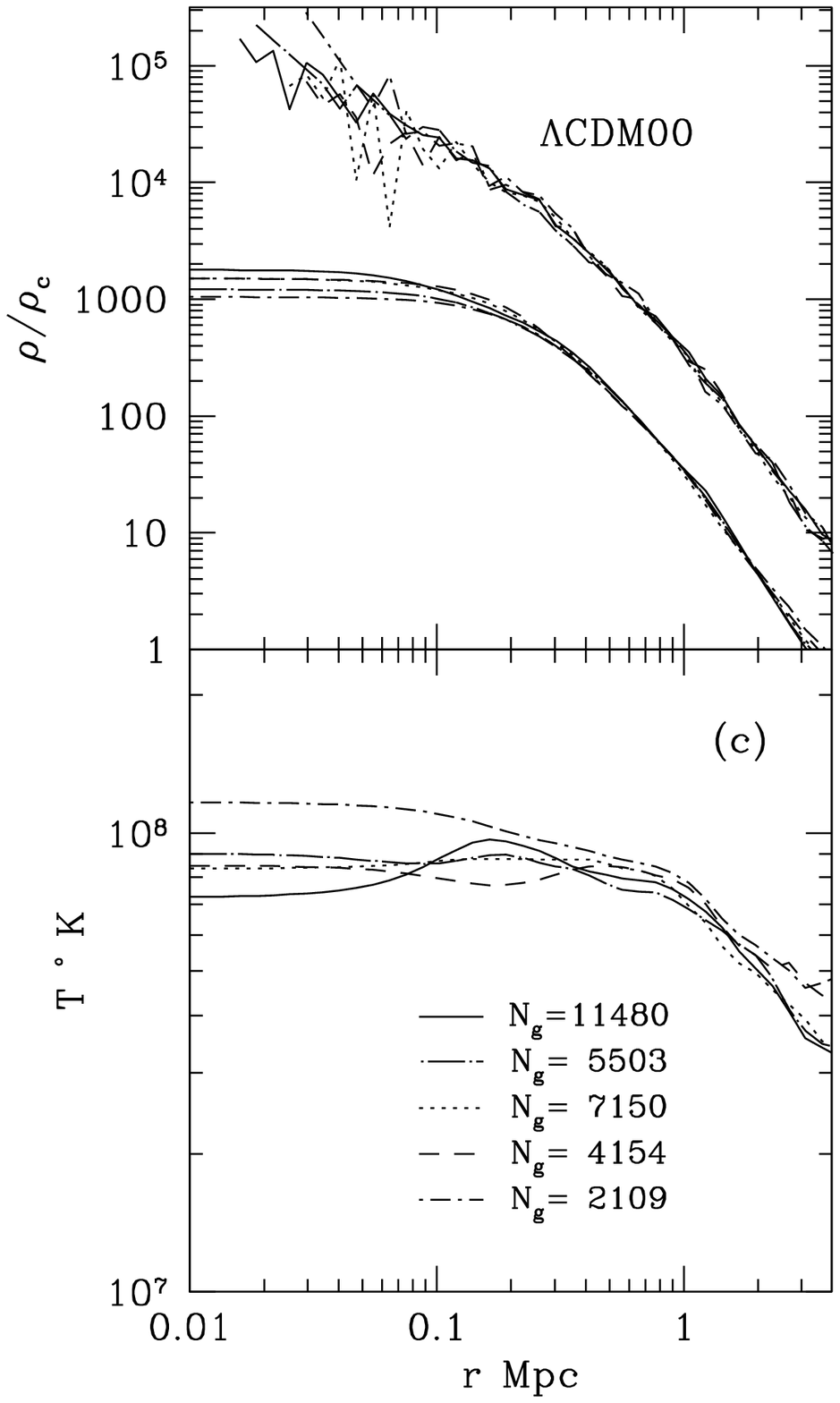,height=11truecm,width=8truecm}%
\psfig{figure=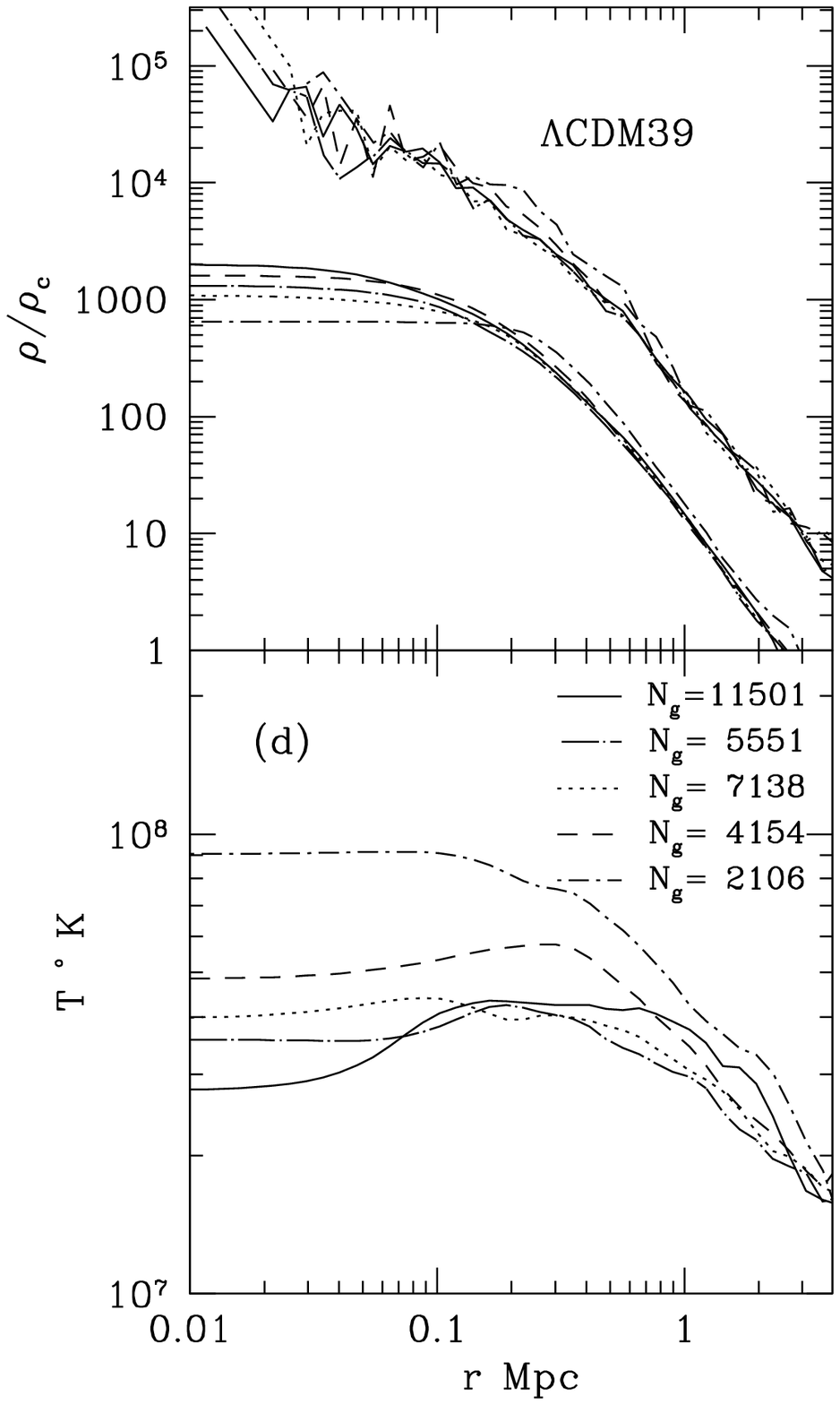,height=11truecm,width=8truecm}%
}}
\vspace*{-0.4cm}
\caption{Density and temperature profile in different numerical tests, for
 the four clusters used as reference. Each panel is for a different
 cluster and in any panel the upper plot is for densities
 and the lower plot for gas temperatures. In each density plot lower
 curves are for the gas and upper curves for the dark matter.
 Different curves are for integrations with a different number of gas 
 particles $N_g$, 
 standard integrations have been performed with $N_g\simeq 5550$. 
The other particle species have their number changed, with respect
the standard value, in proportion to the change in $N_g$.}
\label{fig:ngtst}
\end{figure*}


The 40 cluster models worked out according to the above recipes, for each of
the three models, will be the basis of the statistical discussions of the next
sections. However, in order to assess the reliability of numerical integrations
we have studied how output variables are sensitive to changes in the numerical
parameters, $e.g.$, the number of particles or the softening parameters. 

To this aim, we have taken the most and the least massive clusters found for
the \M~ and \L~ models (labeled $00$ and $39$, respectively) and, for each of
them, we have performed a battery of numerical tests. In Table 2
we report the values of $\varepsilon_g$ and  $N_g$ used in the standard
integrations as well as the values of  $M_{200}$ and $r_{200}$. 

\begin{table}
\label{tab:resol}
\begin{center}
\caption{Reference values for the four 
clusters used in the numerical tests. $N_g$:
number of gas particles, $\varepsilon_g$: value of the 
gas softening parameter in $h^{-1}$ kpc , $M_{200}$: cluster mass within 
$r_{200}$ in $h^{-1} M_{\odot}$, $r_{200}$ in units of $h^{-1}$ Mpc.
These values are for the standard integrations.}
\begin{tabular}{ccccc}
cluster& $N_g$ & $\varepsilon_g$ & $M_{200}$& $r_{200}$ \\ 
\hline
 CHDM00 & 5551 & 50  &  $1.4\cdot 10^{15}$ & 1.83   \\
 CHDM39 & 5575 & 50  &  $4.4\cdot 10^{14}$ & 1.25  \\
 $\Lambda$CDM00 & 5503 & 56  &  $1.2\cdot 10^{15}$ & 1.98   \\
 $\Lambda$CDM39 & 5551 & 42   &  $4\cdot 10^{14}$ & 1.37   \\
\hline
\end{tabular}
\end{center}
\end{table}

The tests start from the same initial conditions, for the TREESPH integrations,
as in the standard case. On the contrary, we vary the gas softening parameter 
$\varepsilon_g$ and/or the number of gas particles $N_g$. In accordance
with them, we also scale the parameters related to different substance 
components. The final part of this section is devoted to a detailed analysis
of the stability of results against such changes. This analysis
allows to conclude that the numerical integrations  are adequately sampled 
for numerical parameters adopted in Table 2.

In Fig. 3 we show the behaviour (at $z=0$) of densities and 
temperatures $vs$ the distance $r$ form the cluster center, for different 
values of $N_g$. In each panel the upper plot is for the gas and dark matter 
densities, the lower plot for the gas temperatures ($T (r)$). Different 
curves refer to $N_g = (2, 1.5, 1, 0.75, 0.5) \times \bar N_g$; here
$\bar N_g$$=5550$ is the standard value of particle number.
For these integrations $\varepsilon_g$ is given in Table 2.

In Fig. 4 we show the same variables of Fig. 3, but
the integrations have been done keeping $N_g$ constant and varying the
softenings ($\varepsilon_g$ and, in proportion, those of the other species). We
considered $\varepsilon_g = (4,2,1,0.5,0.25) \times \bar \varepsilon_g$. Here $
\bar \varepsilon_g$ is the standard values of Table 2.

\begin{figure*}
\centerline{\hbox{%
\psfig{figure=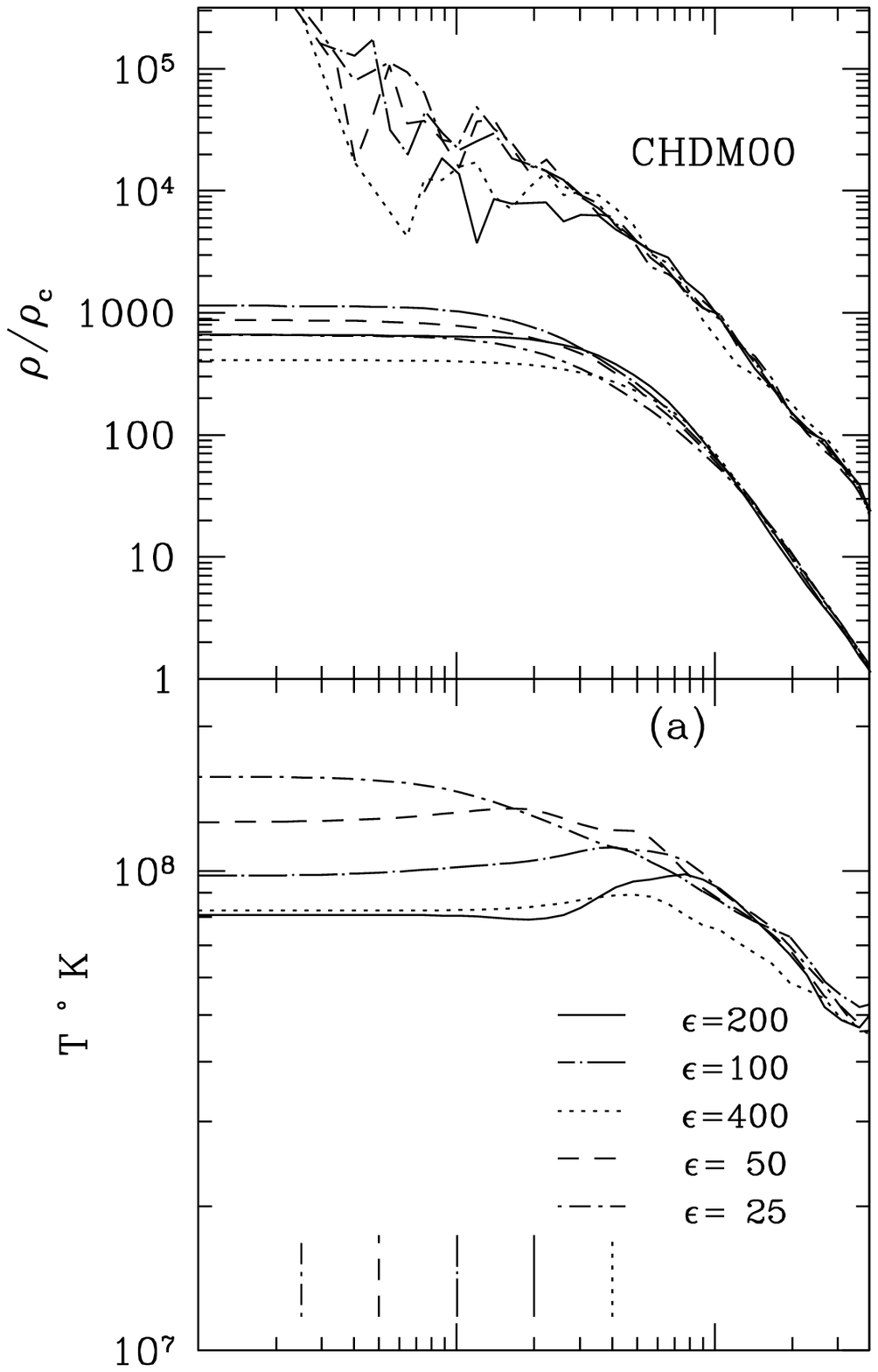,height=11truecm,width=8truecm}%
\psfig{figure=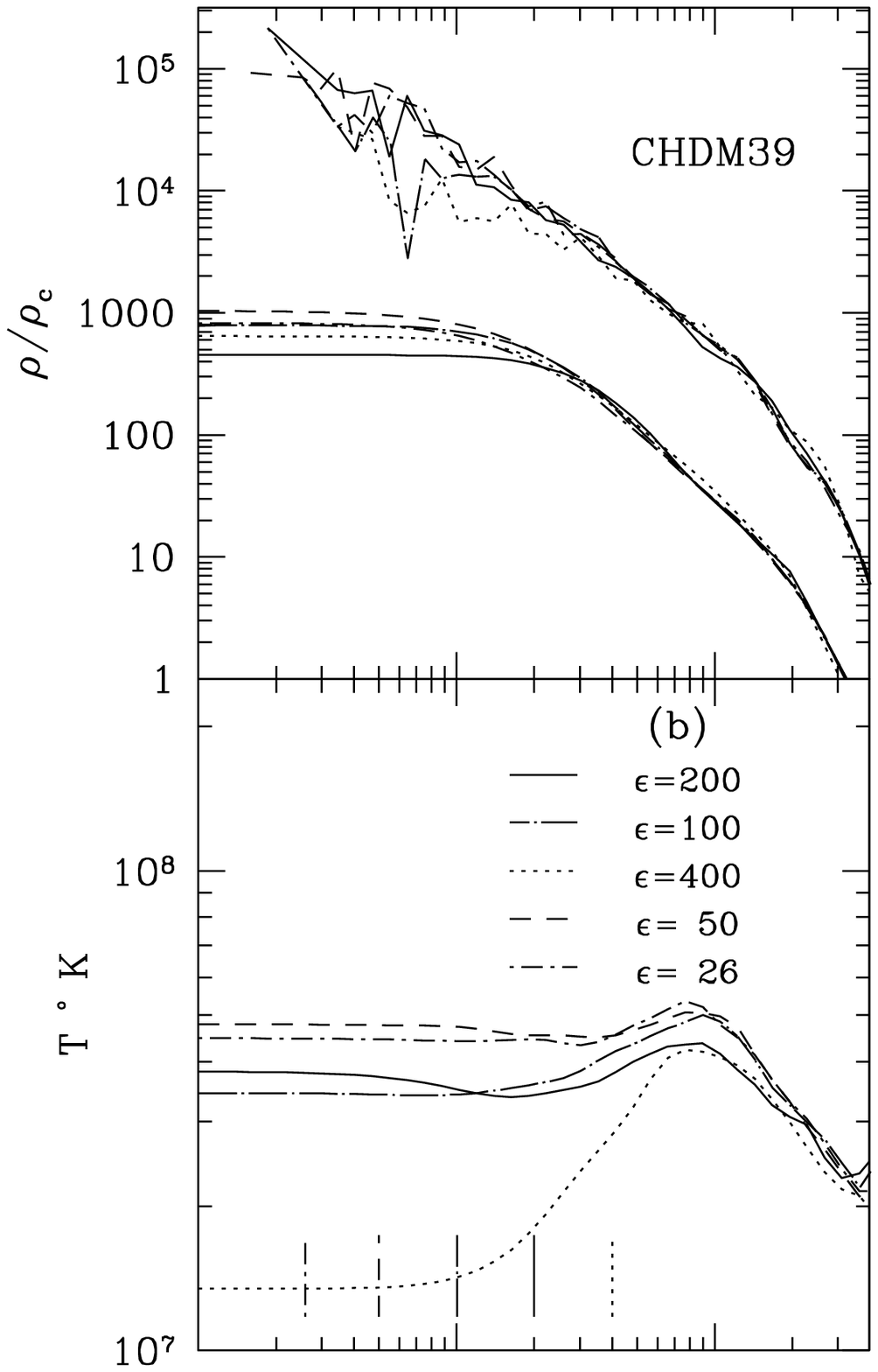,height=11truecm,width=8truecm}%
}}
\centerline{\hbox{%
\psfig{figure=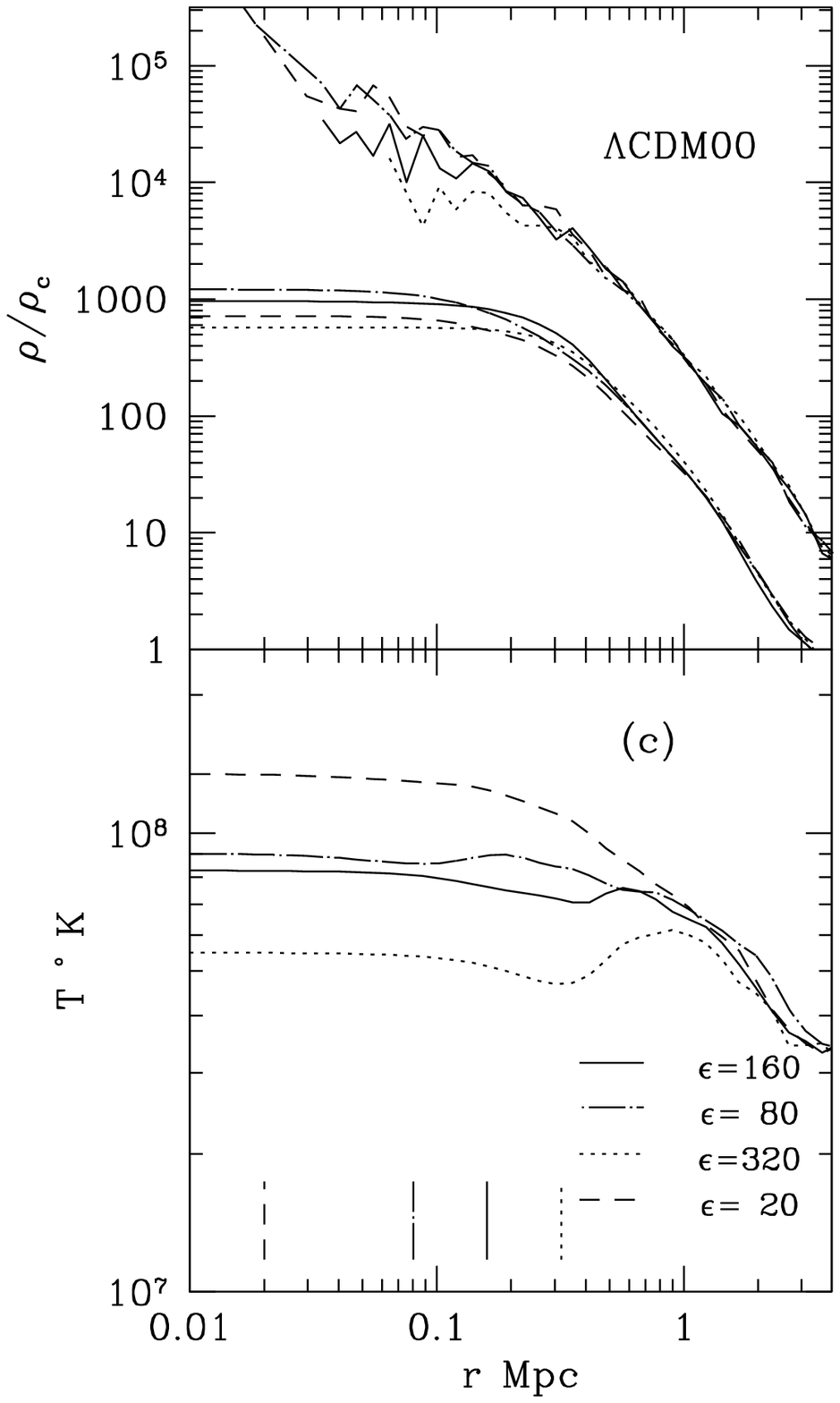,height=11truecm,width=8truecm}%
\psfig{figure=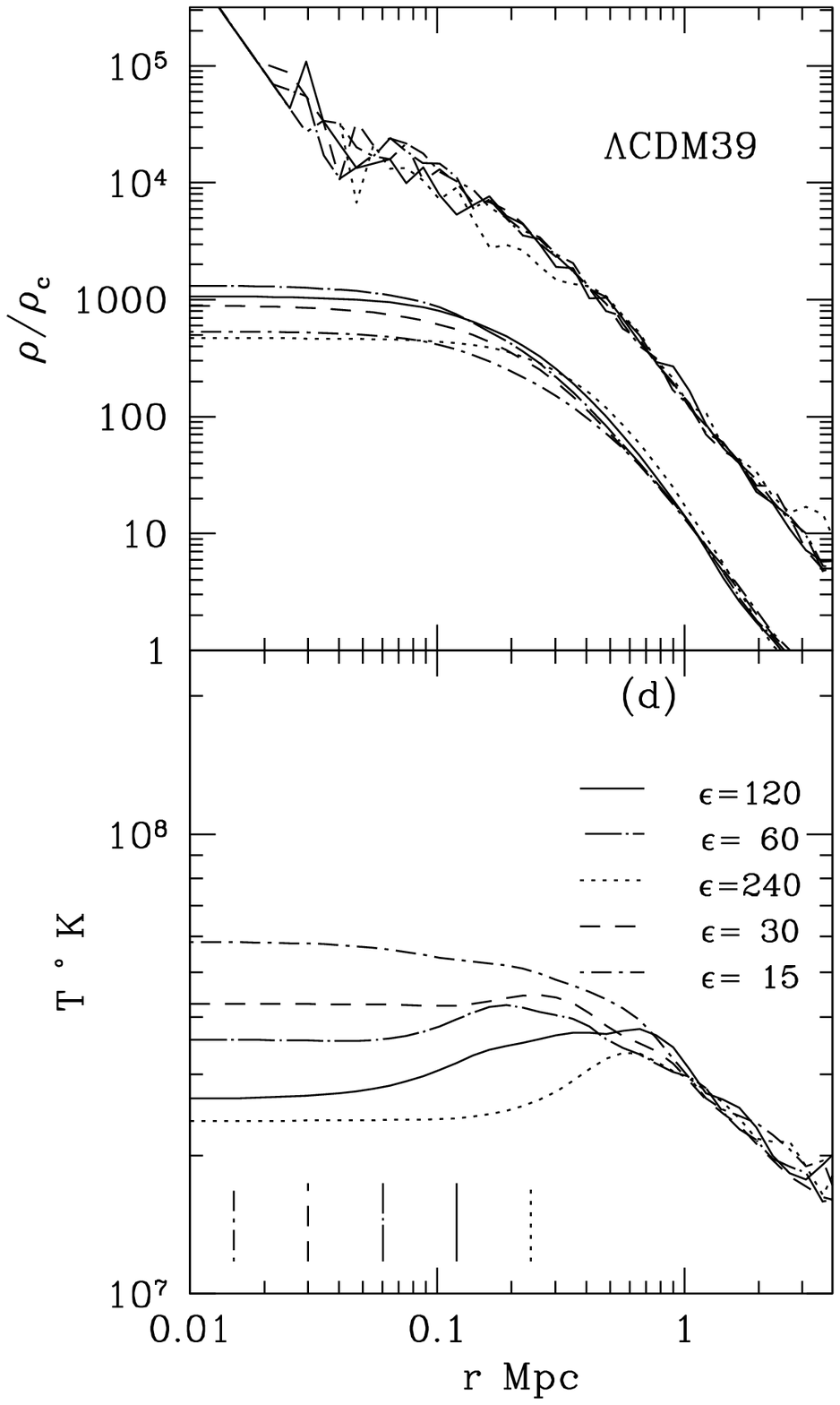,height=11truecm,width=8truecm}%
}}
\caption{The same as in  Fig.~3, but different curves 
are for
different values of $\varepsilon_g$. The values of $\varepsilon_g$ given
in the panels are in kpc, the small vertical lines show their values.
In these integrations the value of $N_g$ is given in Table 2.
}
\label{fig:eptst}
\end{figure*}

Fig.s 3 and 4 show two different behaviours
for $r$ smaller or greater than the core radius $r_c$. At $r > r_c$
discrepancies are modest. However,
Fig.~3 shows that, when $N_g \geq 5,000$, even
the central value of the gas density converges
to within $10\, \%$. Furthermore, regardless of $N_g$, we find that the gas 
core radius $r_c \simeq 0.15$--$0.20\, $Mpc. 
The run of $T(r)$, instead, shows some scatter, at $r \leq r_c$, 
for variations in $N_g$ (\L39 has the strongest variations), but, again,
the central gas temperatures keep within $20\, \%$, when $N_g \geq 5,000$.

There is also a trend for the central values to become smaller when $N_g$ is
greater. This arises because we kept $\varepsilon_g$ constant. Then, when $N_g$
becomes small, two--body heating effects produce an increase in $T(r)$ and the
process is more evident for $r \simlt r_c$. 
In accordance with that, Fig.~4 shows that the central gas 
temperatures tend to be systematically greater when $\varepsilon_g$ decreases. 

In order to keep the characteristic time of the two--body
heating process above the cosmological time, our choice of $\varepsilon_g$ 
turns out to be the best one, for the particle number we took.
In fact, according to Binney $\&$ Tremaine (1987), the 2--body heating time
\begin{equation}
\tau_r = 0.34 \frac {\sigma^3}{G^2 m \rho \ln \Lambda} ~.
\label{eq:tau}
\end{equation}
Here $m$ is the dark particle mass, $\rho$ is the cluster density,
$\sigma$ is the one--dimensional velocity dispersion and $\Lambda
= R/\varepsilon$ gives the Coulomb logarithm associated to the gravitational 
interaction ($R \simeq r_{200}$ ought to be the typical size of the system). 

In CDM models we can estimate $\tau_r$ in a simplified way, but the 
arguments below can be easily translated to other cosmological models.
First of all, according to NFW, the approximate
scaling $\sigma \simeq 911$km sec$^{-1} (M_{200}/10^{15} M_{\odot})^{1/3}$ 
holds. Furthermore, the CDM particle mass $m$ can be expressed as a 
function of the cube side $L_c$ and thus of $M_{200}$. For $N_L=22$ 
it is then easy to find that $m \simeq 2.3 \cdot 10^{-4} M_{200}$.
Accordingly, eq. \ref{eq:tau} yields:
\begin{equation}
\tau_r \simeq 0.7 \cdot 10^6 {\rm Gyr}  
\frac {1}{ ( \rho/\rho_c ) \ln \Lambda }~.
\label{eq:tau2}
\end{equation}

For the most massive cluster we met in CDM (indicated as CDM00),
$M_{200} \simeq 9.3 \cdot 10^{14}  h^{-1} M_{\odot} $, 
 with $r_{200} \simeq 1.5 h^{-1} Mpc$. The heating time decreases
for greater cluster masses and densities. Therefore, this cluster is the
worst possible case we met. 
Furthermore the heating time decreases as the density gets higher, we
evaluate $\tau_r$ at the core radius $r_c \simeq 0.05 r_{200}$, 
approximately the resolution limit 
of our simulations. 
It also well known that, according to NFW, 
the the DM best--fit density profile gives $\rho/ \rho_c =2.2 \cdot 10^4 $,
at $r \simeq r_c$.

As, for CDM00, $\varepsilon=200 \, $kpc and $\ln \Lambda=2.7$,
we shall have the 2--body heating time, for it, is $\tau_r \simeq 12$ Gyr.
This shows that, outside of the core radius, our simulations 
are safely free of 2--body  heating, as is also shown by the way how $T(r)$ 
changes in the numerical tests we performed. 

An inspection of Fig.~4 shows several regularities that
it is worth outlining. First of all, let us take 
as a reference curve $\bar T_g$$(r)$, yielding the $r$ dependence of
the temperature for the softening $\bar \varepsilon_g$.
For $r \simgt r_c$, varying $\varepsilon_g$ has scarce effects on
$T_g$$(r)$. At $r < r_c$ we are approaching the resolution
of our simulations. This is even truer for greater values of
$\varepsilon_g$. However, even for $\varepsilon_g = 400\, $kpc,
the indication of the core structure is mantained.

More in detail, all $T_g(r)$ lie below $\bar T_g$$(r)$, 
when $\varepsilon > \varepsilon_g$ (presumably because of
the reduced spatial resolution of these integrations), while
$T_g(r) > \bar T_g$$(r)$  when $\varepsilon \ll \varepsilon_g$.
But, as previously outlined, for $r > r_c$, we have just quite
a modest shift.

It is also important to outline that the gas densities do not show a 
strong dependence on $\varepsilon_g$, even for $r < r_c$; the lack of 
resolution, therefore, lowers temperatures but scarcely affects
densities. We presume that this is caused because two--body
encounters, when $\varepsilon_g \ll $$\bar \varepsilon_g$,
play a significant role in the core. Here they yield some
depletion of the gas density, close to the center, and a more
relevant increase of its temperature.

Accordingly, the gas density trend as a function of $\varepsilon_g$ seems 
to suggest that $\varepsilon_g \simeq $$\bar \varepsilon_g$ is the optimal 
choice, given the values of the other parameters.

Another way to check the consistency of our results amounts to test
how the overall X--ray cluster luminosity $L_X$ changes in the parameter
space spanned by the numerical tests of Fig.s 3 and 4.
Let us outline that global $L_X$ variations, by themselves,
would not affect power {\sl ratios}. Nevertheless
$L_X$ is quite sensitive to the cluster
central density and a stable $L_X$ output is a key test of the
convergence of gas density and temperature, in this kind of 
numerical simulations.

For the parameter choices used in Fig.s 3 and 4
we evaluated also $L_X$ at $z=0$, according to the expression
given by NFW ($Eq.(6)$). Results are reported
in Fig. 5, where each panel refers to
 a single cluster. Left plots give $L_X$ $vs$ $N_g$ and right
 plots for $L_X$ $vs$ $\varepsilon$. The former ones show that,
within 10$\, \%$, $L_X$ is stable for $ N_g \simgt 5000$
(once again the behaviour of \L39 is peculiar);
$r.h.s.$ plots, instead, deserve a more detailed discussion.

\begin{figure*}
\centerline{\hbox{%
\psfig{figure=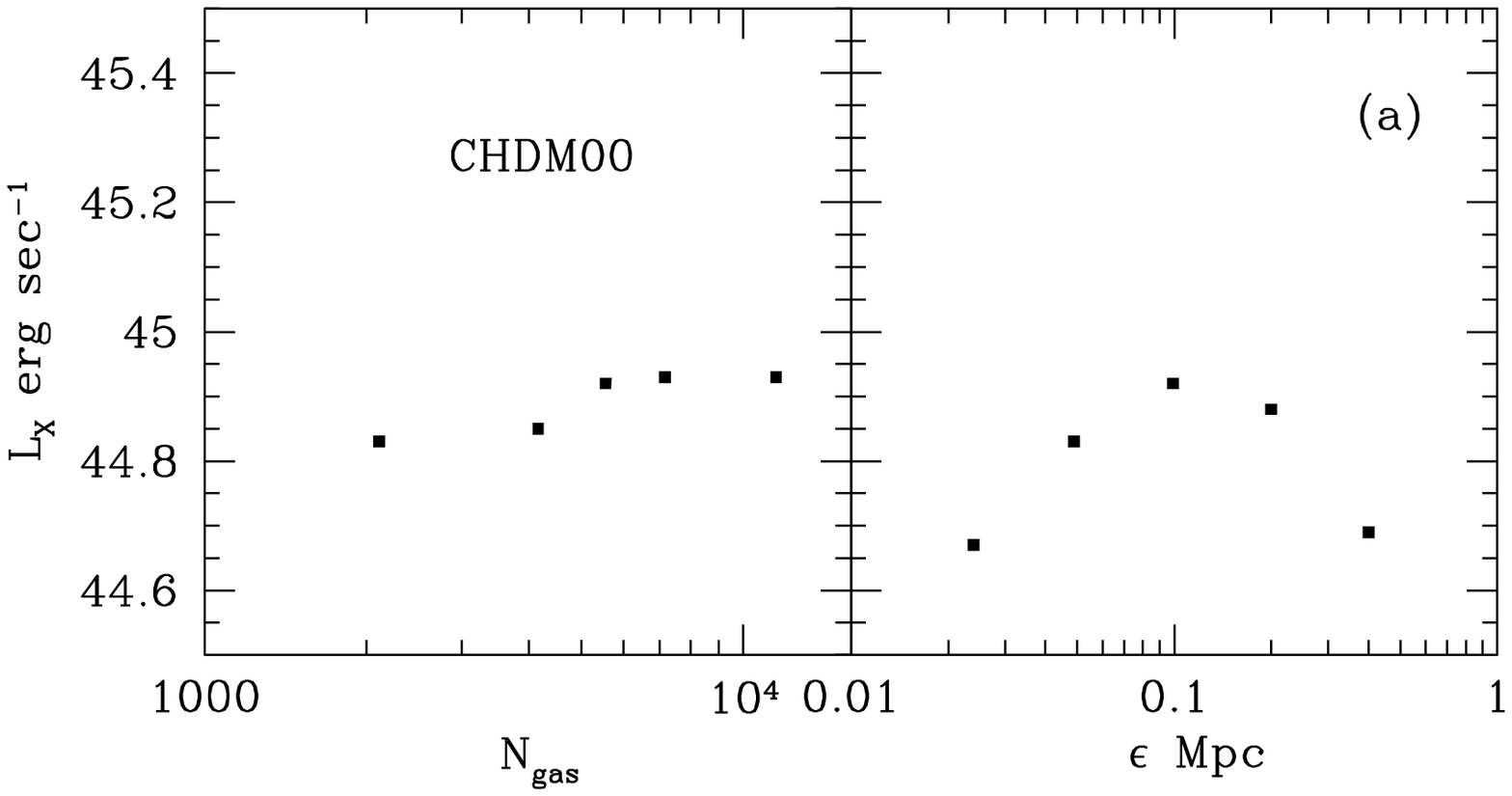,height=5.0truecm,width=11truecm}%
}}
\centerline{\hbox{%
\psfig{figure=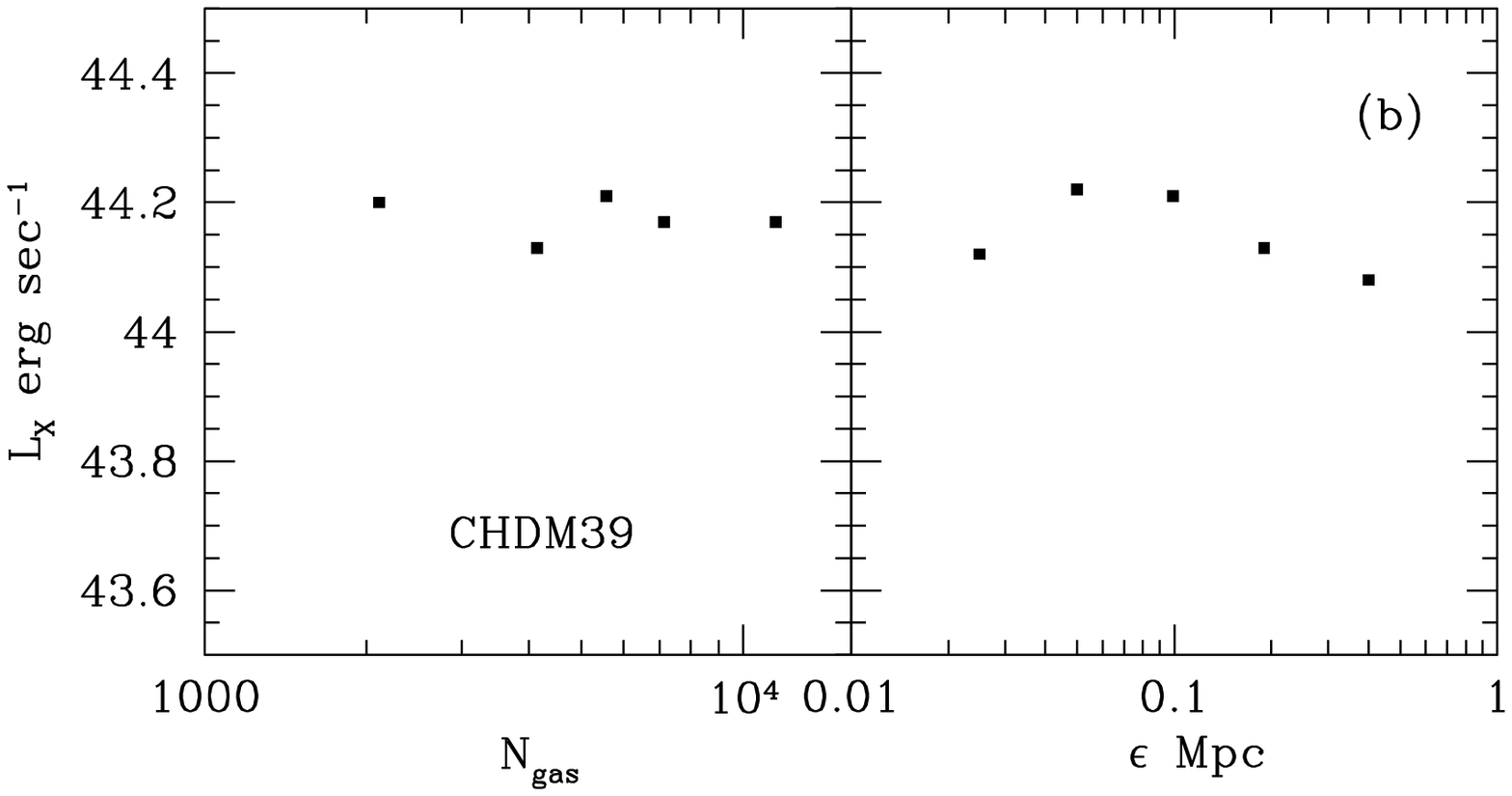,height=5.0truecm,width=11truecm}%
}}
\centerline{\hbox{%
\psfig{figure=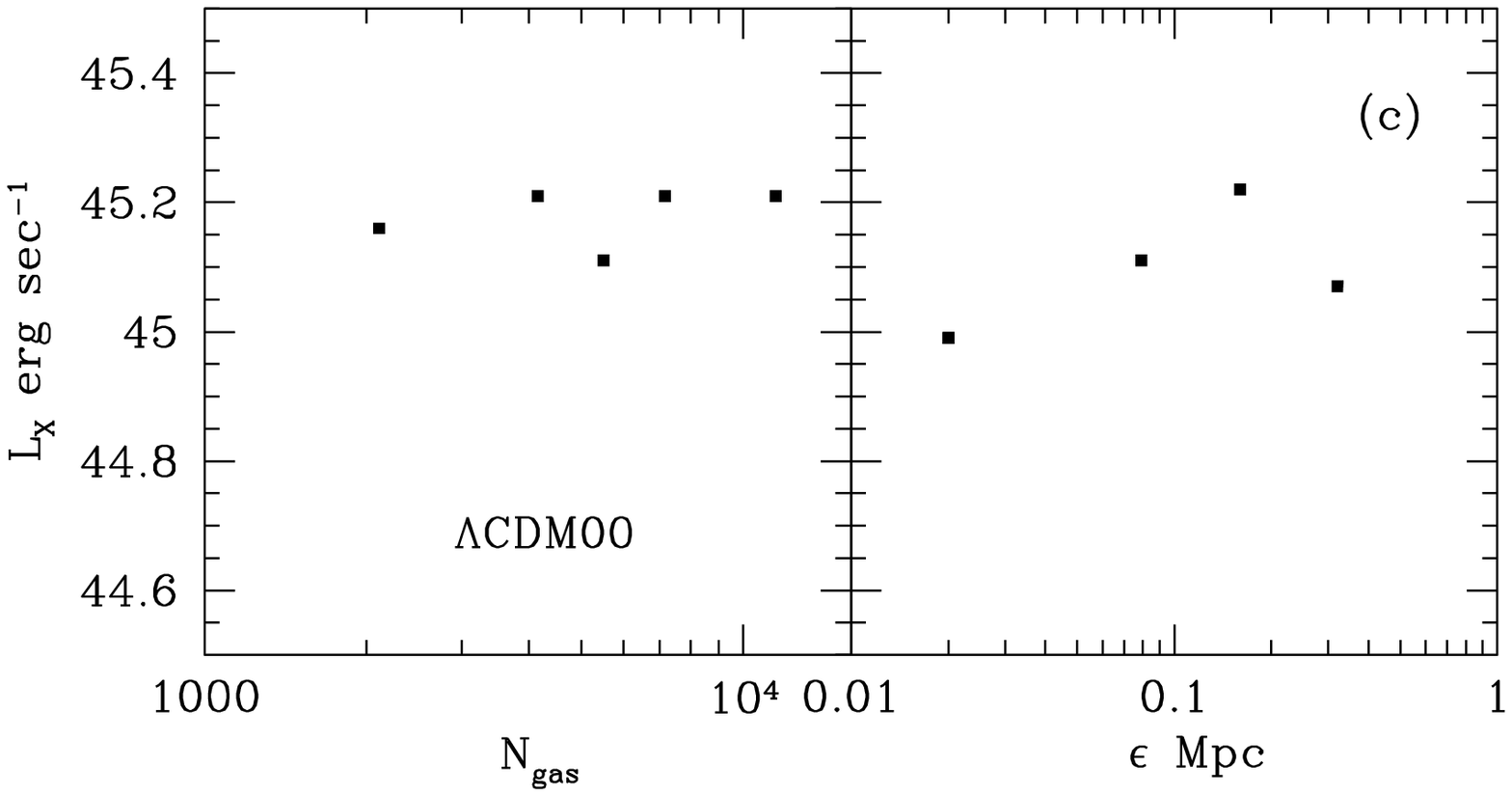,height=5.0truecm,width=11truecm}%
}}
\centerline{\hbox{%
\psfig{figure=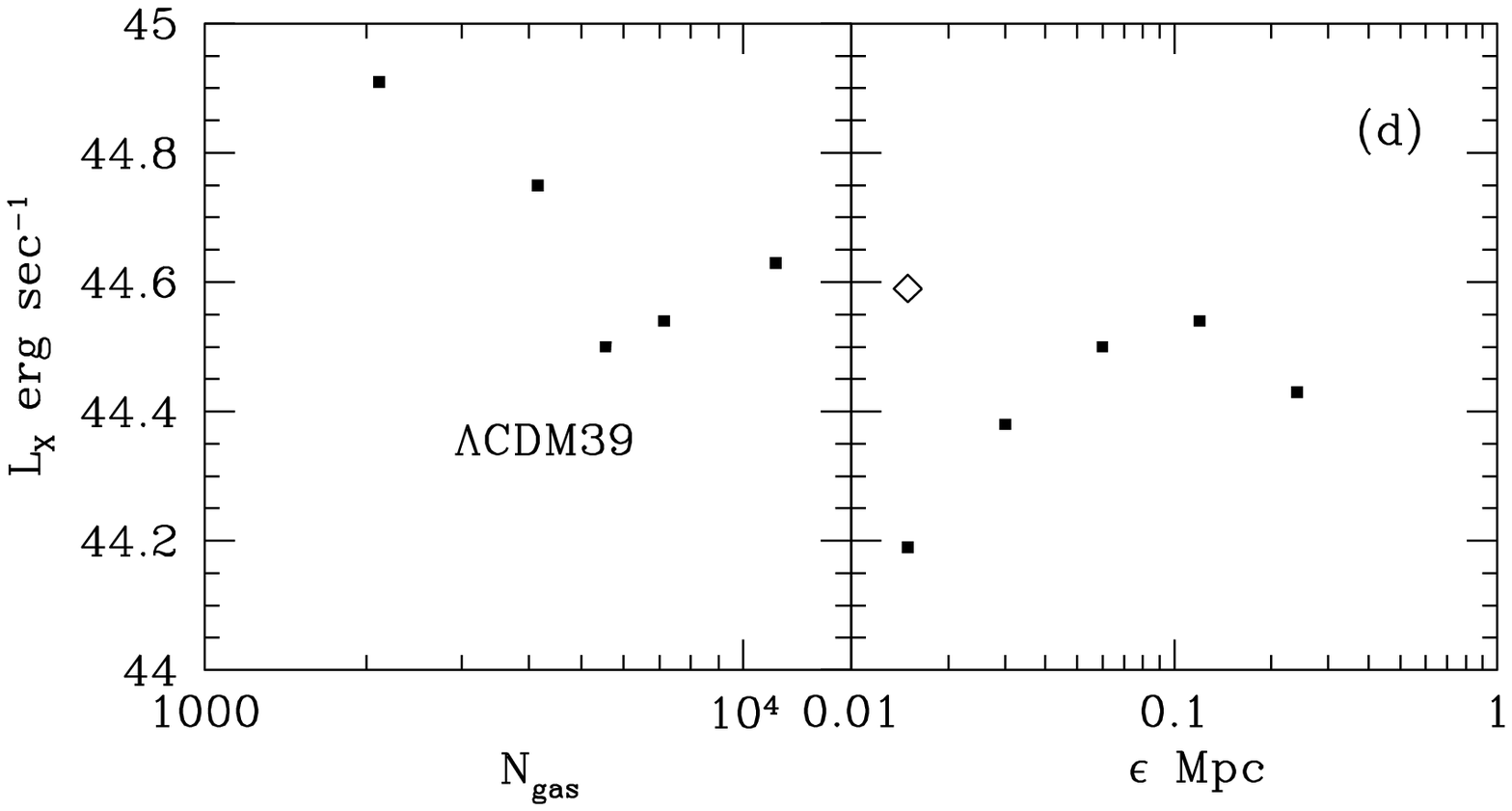,height=5.0truecm,width=11truecm}%
}}
\caption{Final cluster X--ray luminosities for the integrations
of Fig.s 3 and 4. Left plots are for $L_X$
as a function of $N_g$ and $\varepsilon_g$ is constant (Table 2), 
right plots have $N_g\simeq 5550$ and $\varepsilon_g$ varied. The 
open diamond in panel (d) is an integration with 
$\varepsilon_g=15$ kpc and $N_g=11,500$.}
\label{fig:lxtst}
\end{figure*}

Let us again denote  $\bar \varepsilon_g$ the standard values of 
Table 2 and let $\varepsilon_g$ be 
the other softening parameters considered in Fig. 4.

In the $r.h.s.$ plots of Fig. 5, the $L_X$
dependence on $\varepsilon_g$ shows a bell--shaped behaviour, peaked for
$\varepsilon_g \simeq$$\bar \varepsilon_g$.
In fact, for  $\varepsilon_g \gg \bar \varepsilon_g$, the gas spatial
resolution is degraded in the central cluster regions and the
bulk of the contribution to $L_X$ is depressed.
On the contrary, for $\varepsilon_g \ll \bar \varepsilon_g$, two--body heating
effects, as already outlined in the
comments to previous figures, produce a depletion
in $\rho_g(r)$ and an increase of $T_g(r)$ in the central regions.
The net result is a drop in the value of  $L_X$, with respect 
the one given by $\bar \varepsilon_g$.

A countercheck of this interpretation can be obtained by considering
simulations for which $\varepsilon_g $ and $N_g$ are simultaneously
set to $\sim 1/4$ and $\sim 2$ times the  standard values, respectively. 
In this case
the two--body heating time is almost the same as in the standard integration.
We have run such simulation for \L39
and the value found for $L_X$ (reported in the open diamond of 
Fig. 5--d) is indeed close to the standard one.

Let us remind that two-body heating is an effect inversely
proportional to the mass of the most massive particle pairs.
As all clusters, for a given cosmological model, are described 
by the same numbers of particles, indipendently of their total mass, we
might expect stronger effects for top mass clusters.
As Figs. 3, 4 and 5 show, some anomaly was found, instead, for the
cluster \L 39, which is the lightest of the whole ensemble.
It is therefore reasonable to assess
that the scatter seen in previous tests for such cluster does
not arise from two-body heating, but is to be referred to a casually 
irregular distribution. This is confirmed, e.g.,
by the trend of $L_X$ vs. $N_g$, which is anomalous with respect 
to all other clusters (see Fig.~5d left).

We do not expect that relaxation effects can alter significantly 
the values of our estimated \apim~. We have estimated a relaxation time of 
about 12 Gyr at the smallest structure we can resolve ($\simeq 100 Kpc$),
which is about $10\, \%$ of the minimum $R_{ap}$ that we considered when 
the \apim's have been evaluated.
We have however tested how the power ratios 
change when $\varepsilon$ and $N_g$ are varied.
Results of such tests are summarized in Table 3. Here we compare
the variance of \apim's due to the change of line of sight,
with average values and variances of the same \apim's, obtained by
varying softening or particle numbers. In most cases, the latter
variances are smaller than the former one and, in most cases,
the latter average values are consistent with the line--of--sight
average and error. In a large portion of cases, such consistency
occurs at the 1--$\sigma$ level. In three cases (underlined in Table 3),
consistency is not recovered, even at the 3--$\sigma$ level.
Two of them correspond to an anomalously small variance for
changes of the line--of--sight. In the third case, the anomalous
value has also quite a large variance; different selections of
lines--of--sight, in this case, cancel the anomaly; we however refrained
from such $ad$--$hoc$ replacement. All three anomalies concern the 
same cluster, which, once again, is the smallest of its set. However
it is CHDM39, instead of \L 39, which seemed to have a 
suspect behaviour.

Because of the argument oulined above we do not expect two-body effects to
come into play in the evaluation of the power ratios. 
This is confirmed from the results reported in Table 3, where we find
no evidence of any trend in the dependence of \apim\ on
$\varepsilon$ or $N_p$. Altogether, it is licit to conclude
that our results on \apim's are free of biases coming from
two--body relaxation or other resolution effects.

\begin{table}
\begin{small}
\begin{center}
\newsavebox{\foo}
\savebox{\foo}{\parbox{27 truecm}{
\caption{Average values and variances of \apim's obtained varying
line of sight, softening or particle numbers }
\begin{tabular}[t]{l|c|ccc|ccc|ccc} 
\multicolumn{11}{c}{}\\
&&\multicolumn{3}{c}{$\Pi^{(2)} \pm \sigma^{(2)}$} 
&\multicolumn{3}{c}{$\Pi^{(3)} \pm \sigma^{(3)}$} 
&\multicolumn{3}{c}{$\Pi^{(4)} \pm \sigma^{(4)} $} \\
&$R_{ap}$& line of & softening & \# of gas &
line of & softening & \# of gas &line of & softening & \# of gas \\
cluster& & sight & &  particles &sight & &  particles &
 sight &  &  particles \\
\hline \hline
       & 0.4 & 	-4.95 $\pm$ 0.40 & -5.16 $\pm$ 0.05 & -5.37 $\pm$ 0.05 & 
		-7.35 $\pm$ 0.15 & -7.16 $\pm$ 0.17 & -7.12 $\pm$ 0.13 & 
		-7.17 $\pm$ 0.45 & -7.32 $\pm$ 0.07 & -7.58 $\pm$ 0.16 \\
 CHDM00 & 0.8 & -5.76 $\pm$ 0.43 & -5.89 $\pm$ 0.05 & -6.27 $\pm$ 0.08 & 
		-8.08 $\pm$ 0.28 & -8.06 $\pm$ 0.11 & -7.95 $\pm$ 0.07 & 
		-8.59 $\pm$ 1.07 & -8.54 $\pm$ 0.23 & -8.76 $\pm$ 0.09 \\
       & 1.2 & 	-6.47 $\pm$ 0.37 & -6.61 $\pm$ 0.06 & -6.90 $\pm$ 0.04 & 
		-8.96 $\pm$ 0.37 & -8.75 $\pm$ 0.01 & -8.55 $\pm$ 0.04 & 
		-9.31 $\pm$ 0.28 & -9.39 $\pm$ 0.24 & -9.31 $\pm$ 0.02 \\ 
\hline
       & 0.4 & 	-4.77 $\pm$ 0.03 & \underline{ -4.99 $\pm$ 0.05} & 
					\underline{ -5.10 $\pm$ 0.06} & 
		-7.26 $\pm$ 0.26 & -7.21 $\pm$ 0.10 & -7.16 $\pm$ 0.27 & 
		-6.61 $\pm$ 0.07 & \underline{ -7.00 $\pm$ 0.14} & 
					\underline{ -7.25 $\pm$ 0.32} \\
 CHDM39 & 0.8 & -4.89 $\pm$ 0.16 & -5.32 $\pm$ 0.10 & -5.48 $\pm$ 0.18 & 
		-7.04 $\pm$ 0.82 & -7.69 $\pm$ 0.14 & -7.35 $\pm$ 0.29 & 
		-6.55 $\pm$ 0.52 & -7.35 $\pm$ 0.29 & -7.46 $\pm$ 0.46 \\
       & 1.2 & 	-5.45 $\pm$ 0.16 & -5.81 $\pm$ 0.09 & -5.92 $\pm$ 0.17 & 
		-8.85 $\pm$ 0.20 & -8.98 $\pm$ 0.16 & 
					\underline{ -8.18 $\pm$ 0.52} & 
		-7.46 $\pm$ 0.37 & -7.99 $\pm$ 0.06 & -8.04 $\pm$ 0.29 \\
\hline
	      & 0.4 & 	-5.69 $\pm$ 0.15 & -5.36 $\pm$ 0.03 & -5.45 
								$\pm$ 0.03 & 
			-7.55 $\pm$ 0.37 & -7.03 $\pm$ 0.17 & -7.78 
								$\pm$ 0.18 & 
			-8.32 $\pm$ 0.23 & -7.97 $\pm$ 0.43 & -8.47 
								$\pm$ 0.19 \\
$\Lambda$CDM00 & 0.8 & 	-6.31 $\pm$ 0.20 & -5.83 $\pm$ 0.10 & -5.98 
								$\pm$ 0.09 & 
			-8.08 $\pm$ 0.44 & -7.52 $\pm$ 0.12 & -7.73 
								$\pm$ 0.08 & 
			-9.09 $\pm$ 0.40 & -8.47 $\pm$ 0.43 & -8.69 
								$\pm$ 0.25 \\
	      & 1.2 & 	-6.87 $\pm$ 0.28 & -6.37 $\pm$ 0.10 & -6.52 
								$\pm$ 0.08 & 
			-8.30 $\pm$ 0.26 & -8.09 $\pm$ 0.07 & -8.39 
								$\pm$ 0.19 & 
			-9.22 $\pm$ 0.47 & -8.87 $\pm$ 0.37 & -9.28 
								$\pm$ 0.34 \\
\hline
	      & 0.4 & 	-6.20 $\pm$ 0.21 & -6.06 $\pm$ 0.23 & -6.12 
								$\pm$ 0.33 & 
			-8.17 $\pm$ 0.20 & -8.18 $\pm$ 0.63 & -8.15 
								$\pm$ 0.32 & 
			-8.78 $\pm$ 0.46 & -8.69 $\pm$ 0.28 & -8.79 
								$\pm$ 0.36 \\
$\Lambda$CDM39 & 0.8 & 	-6.75 $\pm$ 1.03 & -6.32 $\pm$ 0.28 & -6.56 
								$\pm$ 0.15 & 
			-9.67 $\pm$ 1.76 & -8.25 $\pm$ 0.37 & -8.49 
								$\pm$ 0.59 & 
			-9.45 $\pm$ 1.65 & -8.87 $\pm$ 0.04 & -9.02 
								$\pm$ 0.22 \\
	      & 1.2 & 	-7.10 $\pm$ 1.94 & -6.71 $\pm$ 0.12 & -6.83 
								$\pm$ 0.19 & 
			-9.42 $\pm$ 2.82 & -8.68 $\pm$ 0.04 & -8.49 
								$\pm$ 0.22 & 
			-9.75 $\pm$ 2.93 & -9.56 $\pm$ 0.07 & -9.56 
								$\pm$ 0.70 \\
\multicolumn{11}{c}{}
\label{tab:stab}
\end{tabular}}}
\begin{turn}{90}\usebox{\foo}\end{turn}
\end{center}
\end{small}
\end{table}

\section{Results}
\label{sec:results}

In this work we use power ratios to provide a quantitative characterization for
the global morphology of clusters. In Section \ref{sec:power} we discussed how
they are evaluated on a subsample of ROSAT data. Here we shall work out power
ratios from simulations and discuss how a safe statistical comparison can be
performed between simulations and data. 

Previous work on the same direction has been performed by: BT95, who used just
mock simulations to show that power ratios are an effective tool to distinguish
among different kinds of global morphologies; TB, who used 6 hydro simulations
by NFW for pure CDM, normalized in order that $\sigma_8 = 0.63$, and performed
a comparison with power ratios of ROSAT data (BT96); BX, who used a set of
N--body (non--hydro) simulations for various models, and did a systematic
comparison between them and ROSAT data. 

In this work, we use a large set of hydro simulation (120), for 3 different
cosmological models, to perform a similar comparison. 
The need to widen
the extension of the sample had been felt also by BX,
who tried to satisfy it by using non--hydro simulations, although
they are quite aware of the problems this implies.
In full agreement with their reserves, we find that the use
of non--hydro simulations, for this kind of problems,
can be misleading. This conclusion follows a detailed comparison
of results concerning the global morphology of the same clusters,
using their gas and dark matter distributions.

Henceforth, previous work on power ratios was essential
to show that this is an effective tool to study the global properties
of clusters and to discriminate, through them, among cosmological models.
In this work, instead, we are able to provide the first
set of simulations adequate to make use of such tool and to
try a to select the {\sl best} cosmological model.
In this context, a basic contribution concerns the use
of statistical tools, which are discussed in this section.

\subsection{Power ratios for different cosmological models}
\label{sub:ratios}

In this sub--section we report the power ratios \apim\ that we computed for our
three cosmological models from the gas distributions. The average and
dispersions of the \apim\ are given in Table 4, together with
the corresponding ROSAT values. The procedure followed to evaluate them was
outlined in section \ref{sec:power}. They were obtained using a single sequence
of random planes and averaging over different realization for the redshift
distribution. The three different apertures are in units of $h^{-1}$Mpc and
correspond to those given by BX in their Table 2. 

\begin{table}
\label{tab:avedisp}
\begin{center}
\caption{Averages and variances of \apim\ for ROSAT and simulated 
clusters in the three cosmological models.}
\begin{tabular}{l|c|c|c}
&\multicolumn{3}{c}{$ \Pi^{(2)} \pm \sigma^{(2)}$}   \\
 &{0.4 Mpc/h} & {0.8 Mpc/h} & {1.2 Mpc/h}
\\ 
\hline
 ROSAT & -5.74 $\pm$  0.53  & -5.96 $\pm$ 0.66 & -6.46 $\pm$ 0.81 \\
 CDM  &  -5.47 $\pm$  0.54  & -6.17 $\pm$ 0.88  & -6.12 $\pm$ 0.85 \\
 $\Lambda$CDM  & -5.63 $\pm$ 0.45 &  -6.36 $\pm$ 0.73  &   -6.79 $\pm$ 0.86 \\
 CHDM  & -5.40 $\pm$ 0.54 &  -6.16 $\pm$ 0.71  &   -6.63 $\pm$ 0.72\\
\multicolumn{4}{c}{} \\
& \multicolumn{3}{c}{$\Pi^{(3)} \pm \sigma^{(3)}$} \\
 &{0.4 Mpc/h}&{0.8 Mpc/h}&{1.2 Mpc/h}
 \\
\hline
 ROSAT & -7.43 $\pm$ 0.77&  -7.39 $\pm$ 0.73  &  -7.75  $\pm$ 0.72 \\
 CDM  &  -7.16 $\pm$ 0.73 & -7.55 $\pm$  0.93 &  -7.48  $\pm$ 0.93 \\
 $\Lambda$CDM  & -7.66 $\pm$ 0.83 &  -8.33 $\pm$ 1.07 & -8.67 $\pm$ 1.17 \\
 CHDM  & -7.07 $\pm$ 0.79 &  -7.68 $\pm$ 0.85  &  -7.96 $\pm$ 0.80 \\
\multicolumn{4}{c}{} \\
 &\multicolumn{3}{c}{$ \Pi^{(4)}  \pm \sigma^{(4)}  $}   \\
 &{0.4 Mpc/h}&{0.8 Mpc/h}&{1.2 Mpc/h}
 \\
\hline
 ROSAT & -7.90 $\pm$ 0.63 &  -7.90 $\pm$ 0.84 &  -8.10 $\pm$ 0.83 \\
 CDM  & -7.69 $\pm$ 0.69 &  -8.21 $\pm$ 1.10 &  -7.98 $\pm$ 1.07 \\
 $\Lambda$CDM  & -8.02 $\pm$ 0.73 &  -8.82 $\pm$ 1.08 &  -9.22 $\pm$ 1.21 \\
 CHDM  & -7.56 $\pm$ 0.84 &  -8.31 $\pm$ 0.94 &  -8.60 $\pm$ 0.89 \\
\end{tabular}
\end{center}
\end{table}

In fig.~6 we report the \apim\ distribution for ths \M~ model.
Similar plots could be given for the other cosmological models.
Heavy lines give the model cluster outputs, while light lines
correspond to ROSAT cluster data. Each line correspond to the
same $R_{ap}$, each column to the same \apim. Such figure
shows a basic agreement between model and data clusters, which
can also be seen through the figures reported in Table 4.

\begin{figure*}
\label{fig:histom}
\centerline{\mbox{\epsfysize=16.0truecm\epsffile{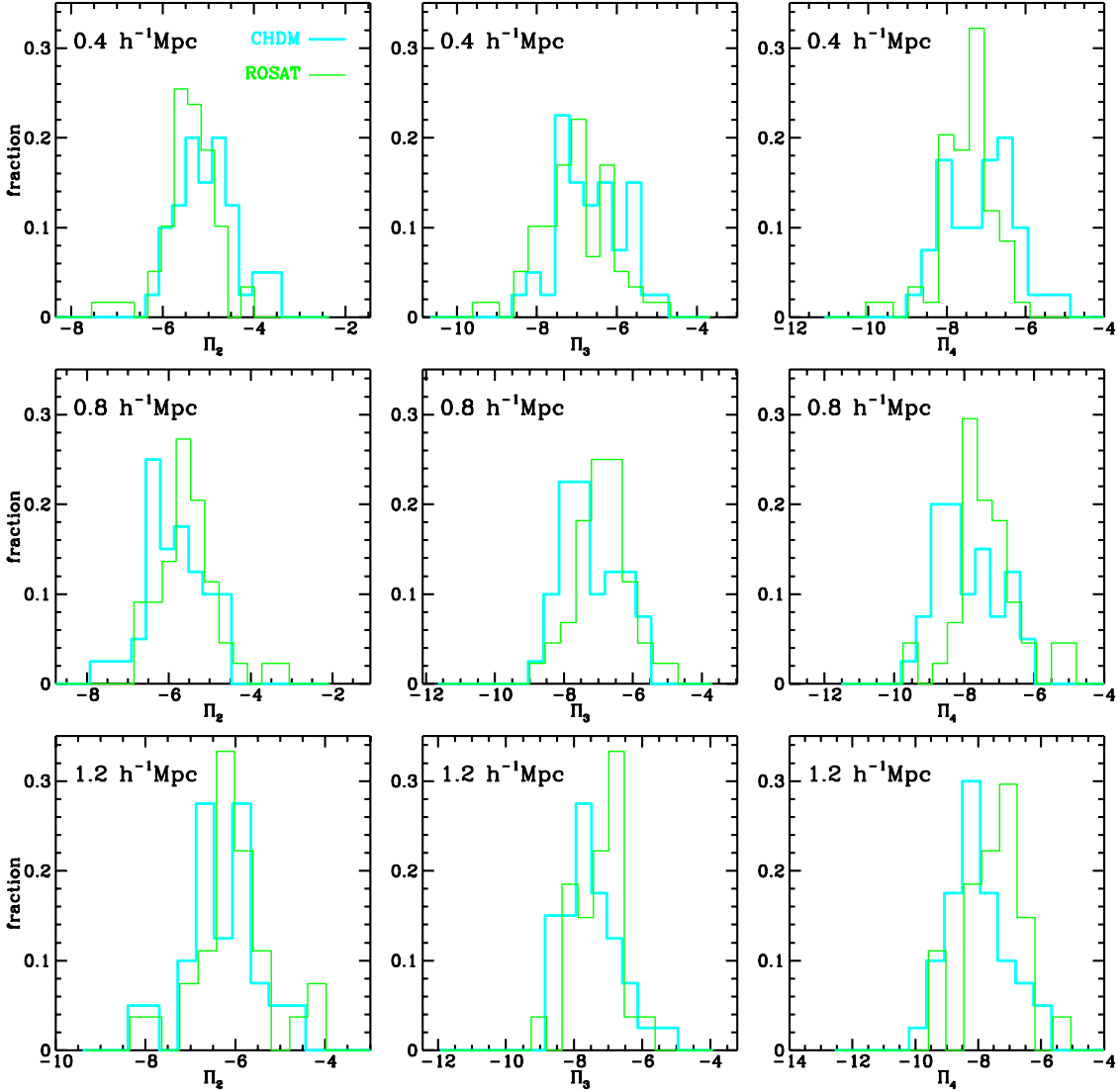}}}
\caption{Histograms of $\Pi^2, \Pi^3, \Pi^4$ in the $R_{ap}=0.4,0.8,1.2
h^{-1}$Mpc for CHDM clusters.}
\end{figure*}

More in detail, for
$R_{ap} h / {\rm Mpc} = 0.4$ and for all moments, ROSAT and simulation
distributions mostly agree, regardless of the cosmological model. This may be
an indication that, over small scales, initial conditions are mostly erased by
non--linear dynamics and a relaxation regime, independent of the model, is
attained. For $R_{ap} h /{\rm Mpc} = 0.8$ and 1.2, instead, all simulated
distributions are shifted towards smaller values. However, while CDM and \M~
are still marginally consistent with data, \L~ is far below them. This is
particularly relevant for $\Pi^{(2)}$, but also $\Pi^{(3)}$ and $\Pi^{(4)}$
show a similar trend. 

In order to quantify these differences, we used 
the Student t--test, the F--test and the
Kolgomorov--Smirnov (KS) test. Let us recall that: (i) The Student t--test
allows to compare two data sets with different means and finds the probability
$p$--$t$ (reported in Table 5) that, owing to the distribution of
the outputs about such means, the two sets are originated by the same process.
(ii) The F--test, instead, compares two data sets with different variances and
finds the probability $p$--$F$ (reported in Table 5) that the two
sets are originated by the same process. (iii) Finally, the KS--test also
compares two data sets with different distributions and finds the probability
$p$--$KS$ (reported in Table 5) that such different distributions
can arise from the same process. 

\begin{table*}
\label{tab:test}
\begin{center}
\caption{Statistical tests applied to the power ratio distribution of ROSAT 
and simulated clusters: $p_t$ is the Student t--test applied to the means,
$p_F$ is the F--test for variances and $p_{KS}$ is the KS statistics to
discriminate two distribution.}
\begin{tiny}
\begin{tabular}{l|ccc|ccc|ccc}
\multicolumn{10}{c}{}\\
\multicolumn{10}{c}{$\Pi^{(2)}$}  \\
\multicolumn{10}{c}{} \\
 &\multicolumn{3}{c}{0.4 Mpc/h}&\multicolumn{3}{c}{0.8 Mpc/h}&
 \multicolumn{3}{c}{1.2 Mpc/h}
 \\ 
models & $p_t$ & $p_F$ & $p_{KS}$ &$p_t$&$p_F$&$p_{KS}$ & 
$p_t$ & $p_F$ & $p_{KS}$ \\ \hline
ROSAT-CDM   &   .16E--01 &  .86E+00 & .37E--01  & .21E+00 & .62E--01 & .14E+00 
& .11E+00 &  .80E+00 &  .99E--01 \\
ROSAT-$\Lambda$CDM  &   .29E+00 &  .27E+00 &  .67E+00 &  .87E--02 & 
.49E+00 &  .31E--02   & .12E+00 &  .75E+00 &  .11E--01 \\
ROSAT-CHDM &  .32E--02 &  .88E+00  & .85E--02 & .17E+00 & .60E+00 & .24E+00 &
 .37E+00 & .49E+00 &  .52E+00 \\
CDM -$\Lambda$CDM &  .16E+00 &  .24E+00  & .26E+00 &   .29E+00 &  .25E+00 &  
.16E+00 & .78E--03 &   .94E+00 &  .24E--03 \\
CDM -CHDM  &   .60E+00 &  .98E+00 &  .91E+00 & .98E+00 &  .19E+00 &  
.57E+00  &  .48E--02 &  .30E+00 &  .55E--01 \\
$\Lambda$CDM -CHDM  &  .49E--01 &  .24E+00 &  .97E--01 &   .22E+00 &  .87E+00 
&  .40E+00 &   .38E+00 &  .26E+00 &  .26E+00  \\
\multicolumn{10}{c}{} \\
\multicolumn{10}{c}{} \\
\multicolumn{10}{c}{$\Pi^{(3)}$}\\
\multicolumn{10}{c}{} \\
 &\multicolumn{3}{c}{0.4 Mpc/h}&\multicolumn{3}{c}{0.8 Mpc/h}&
 \multicolumn{3}{c}{1.2 Mpc/h}
 \\ 
models & $p_t$ & $p_F$ & $p_{KS}$ &$p_t$&$p_F$&$p_{KS}$ & 
$p_t$ & $p_F$ & $p_{KS}$ \\ \hline
ROSAT-CDM  &  .95E--01 &  .69E+00 &  .41E+00 &   .37E+00 &  .11E+00 &  
.18E+00   &  .20E+00 &  .16E+00 &  .11E+00 \\
ROSAT-$\Lambda$CDM  &  .15E+00 &  .62E+00 &  .32E+00 &   .15E--04 &  .15E--01 
&  .33E--05   & .17E--03  & .11E--01  & .28E--02 \\
ROSAT-CHDM  &  .29E--01 &  .86E+00 &  .13E+00 &   .97E--01 &  .31E+00 &  
.64E--02   & .27E+00  & .57E+00  & .36E+00 \\
CDM -$\Lambda$CDM   &   .53E--02 &  .41E+00 &  .15E--01 &  .90E--03 &  .40E+00 
&  .72E--02   & .28E--05  & .17E+00 &  .24E--03 \\
CDM -CHDM  &   .58E+00 &  .59E+00 &  .91E+00 &   .53E+00 &  .58E+00 & 
.57E+00     & .15E--01 &  .34E+00 &  .55E--01 \\
$\Lambda$CDM-CHDM   &   .16E--02 &  .78E+00 &  .15E--01 &  .36E--02 &  .17E+00 
&  .72E--02    & .23E--02 &  .21E--01  & .15E--01 \\
\multicolumn{10}{c}{}\\
\multicolumn{10}{c}{}\\
\multicolumn{10}{c}{ $\Pi^{(4)}$}\\
\multicolumn{10}{c}{}\\
 &\multicolumn{3}{c}{0.4 Mpc/h}&\multicolumn{3}{c}{0.8 Mpc/h}&
 \multicolumn{3}{c}{1.2 Mpc/h}
 \\ 
models & $p_t$ & $p_F$ & $p_{KS}$ &$p_t$&$p_F$&$p_{KS}$ & 
$p_t$ & $p_F$ & $p_{KS}$ \\ \hline
ROSAT-CDM  & .13E+00 &  .57E+00 &  .71E--01 &   .14E+00 &  .86E--01 & .39E--01 &
.60E+00 &  .17E+00 &  .45E+00 \\
ROSAT-$\Lambda$CDM  &    .38E+00 &  .31E+00 &  .32E--01 &   .32E--04 &  
.11E+00 &  .40E--06 & .28E--04  & .45E--01  & .28E--04 \\
ROSAT-CHDM &   .25E--01  & .55E--01  & .11E--01 &   .37E--01 &  .47E+00 &  
.33E--02 &   .25E--01 &  .71E+00 &  .20E--01 \\
CDM -$\Lambda$CDM  &  .44E--01 &  .69E+00 &  .16E+00  &  .15E--01 &  .91E+00  
& .72E--02    & .52E--05 &  .45E+00 &  .33E--04 \\
CDM-CHDM  &   .44E+00 &  .23E+00 &  .76E+00  &  .68E+00 &  .33E+00 &  .57E+00
 &  .59E--02 &  .25E+00 &  .29E--01 \\
$\Lambda$CDM-CHDM  &   .11E--01 &  .42E+00 &  .29E--01  &  .26E--01 &  .39E+00 
&  .72E--02    & .98E--02 &  .58E--01 &  .15E--01 \\
\end{tabular}
\end{tiny}
\end{center}
\end{table*}

Even apart from the smallest aperture, which seems not so significant,
according to Table 5 $p$--$t$ ($p$--$F$, $p$--$KS$) is roughly in
the ranges 0.11--0.60 (0.11--0.80, 0.04--0.45) for CDM, 0.03--0.37 (0.47--0.71,
0.33$\cdot 10^{-2}$--0.52) for \M, 0.15$\cdot 10^{-4}$--0.87$\cdot 10^{-2}$
(0.01--0.75, 0.4$\cdot 10^{-6}$--0.01) for \L. Let us recall that only figures
below $5\, \%$ can be taken at face--value. Greater values are to be considered
underestimates of the consistency probabilities. 

Such figures seem to exclude that \L\ can be considered a reasonable
approximation to data. The best score belongs to CDM, but also \M\ is not fully
excluded and different mixtures could certainly have better performance. 

Table 5 provides also comparisons among different models. CDM and
\M, again, do not show a marked disagreement. \L\ is significantly different,
as follows also from the figures reported hereabove. A possible interpretation
of such output is that the actual amount of substructures is  governed by
$\Omega_0$ rather than by the shape of power spectra. An inspection of the
model clusters actually shows that the \L~ model does produce less
substructures than the other models do. 

As a further test we have considered a two--dimensional generalization
of the KS--test. The test applies to two 2--dimensional distributions
and gives the probability that they originate 
from the same process (Peacock 1983, Fasano and Franceschini 1987,
Press \etal 1992).
The results (see Table 6) confirm the KS test and show that 
\apim\ values, for the smallest aperture $R_{ap} = 0.4\, h^{-1}$Mpc,
are not so discriminatory. On the contrary, at greater apertures,
tests on \L~ often yield probabilities of a few thousands. \M, instead,
is almost anywhere consistent with data, at least at the $\sim$2--$\sigma$
level, apart of a case, where, however, the probability is not
much lower. CDM, perhaps, has a slightly worse performance, as
it never overcomes a 10$\, \%$ probability and is out of $\sim$2--$\sigma$'s
in 2 cases.

\begin{table}
\label{tab:pf2}
\begin{center}
\caption{Probabilities worked out using the PF2--test, a 2--dimensional 
generalization of KS--test. Notice that, for \L~ models, they may
be as low as a few parts on a thousand; the greatest values are met
for CHDM, which, apart of a case, always agrees with data at least at
2--$\sigma$ level; the performance of CDM is only slightly worse.}
\begin{tabular}{c |c| c| c| c}
&$R_{ap} h/{\rm Mpc}$& CDM &$\Lambda$CDM&CHDM\\ 
\hline
$\Pi_2-\Pi_3$ & 0.4 & 6.25$\, \%$ & 10.57$\, \%$ & 11.73$\, \%$\\
$\Pi_2-\Pi_4$ & 0.4 & 5.44$\, \%$ & 17.26$\, \%$ &  4.89$\, \%$\\
$\Pi_3-\Pi_4$ & 0.4 & 8.03$\, \%$ &  8.35$\, \%$ &  6.83$\, \%$\\
$\Pi_2-\Pi_3$ & 0.8 & 6.66$\, \%$ &  0.71$\, \%$ &  3.45$\, \%$\\
$\Pi_2-\Pi_4$ & 0.8 & 9.25$\, \%$ &  1.81$\, \%$ &  5.42$\, \%$\\
$\Pi_3-\Pi_4$ & 0.8 & 2.79$\, \%$ &  0.13$\, \%$ &  1.41$\, \%$\\
$\Pi_2-\Pi_3$ & 1.2 & 4.35$\, \%$ &  2.46$\, \%$ & 10.21$\, \%$\\
$\Pi_2-\Pi_4$ & 1.2 & 3.67$\, \%$ &  0.35$\, \%$ &  8.82$\, \%$\\
$\Pi_3-\Pi_4$ & 1.2 & 6.62$\, \%$ &  0.20$\, \%$ & 11.60$\, \%$\\
\hline
\end{tabular}
\end{center}
\end{table}

\subsection{Power ratio evolution}
\label{sub:evolution}

During its evolution, a cluster moves along a line ({\sl evolutionary 
track}, see BT96 and TB) in the 3--dimensional space spanned by the 
\apim's. Starting from
a configuration away from the origin, corresponding to a large amount
of internal structure, it evolves towards isotropization and 
homogeneization. This motion does not occur with a steady trend,
and bursts of structure appear when further matter lumps
approach the cluster potential well, to be absorbed by it.
However, the evolutionary track eventually approaches the origin,
and this can be more easily appreciated by averaging over the
contributions of several clusters.

Actual data, of course, do not show the motion of a single cluster
along the evolutionary track. Different clusters, however, lie at
different redshifts and, in average, can be expected to
describe a succession of evolutionary moments.

In Fig. 7 we give 9 plots for the space containing the
evolutionary tracks projected on 3 planes and at three different $R$'s.
For the sake of example, in Fig. 7 we describe the
behaviour of \M. Filled dots refer to a single selection of simulated 
clusters. Their distributions show a linear trend and a linear regression 
gives place to the straight lines reported in the plots.
Crosses, instead, indicate the location of data clusters.
Their distributions do not show such a trend as simulated points
and are more scattered than \M\ clusters.

A quantitative way 
to compare the correlations of two 2--dimensional distributions on a plane 
$x,y$, starts from defining the averages $\bar x$, $\bar y$ of $x_i$ and 
$y_i$ values and the deviations $\xi_i = x_i - \bar x$ 
and $\eta_i = y_i - \bar y$. Then the linear correlation coefficients read
\begin{equation}
r = (\Sigma_i \xi_i \eta_i) / (\sqrt{\Sigma_i \xi_i^2} 
\sqrt{\Sigma_i \eta_i^2}) ~.
\end{equation}
The use of this tool indicates that the probability of obtaining such 
different correlation coefficients is quite small and would seem to 
exclude the validity of all models considered.

There may be a number of reasons for such discrepant behaviours.
The time elapsed from $z = 0.2$ up to now is approximately
a quarter of the life of the Universe. During such time, galaxies
evolve and may acquire different relevance in the overall $X$--ray
emission, the intracluster gas is suitably enriched and/or undergoes
complex cooling processes, environmental
effects and other physical variables may have had suitable
trends. When evolution is not our aim, measures on
samples with similarly weighted contributions from different redshift
ranges may keep full significance, while, on the contrary,
separating contributions from different $z$ values 
might be demanding too much. We should therefore refrain from
concluding that none of the models inspected fits the (un--)observed 
evolutionary trend. We rather suggest that the evolutionary trend in data 
is (at least partially) hidden by other effects.

We seeked a confirm of the latter hypothesis by trying to use a softer 
statistical approach to compare the evolution shown by simulations
with data. The slope of the straight line in Fig.~7, 
obtained through a linear
regression, is, by definition, a suitably weighted average among 
the slopes of straight lines through any couple of points.
Accordingly, a linear trend is visible as a peak in the slope distribution.

If a similar trend is not completely hidden in data points, a
peak should appear also in the distribution of the slopes of lines
throught pairs of data points.
In Fig.~8 we compare the distributions of
slopes for data points (thin histograms) and for simulated points (thick 
histograms)
for CDM and CHDM models. 
This is done starting from the points shown in Fig. 7 
for \M~ and for a similar selections for CDM. Fig.s 8, as expected,
show a set of peaks for model clusters, 
more relevant for greater aperture radii.

Let us however outline that some peak structure is also present in data points.
Also an eye comparison of CDM and \M~ with data shows that \M~ is
substantially favoured. The only plot showing an opposite trend is $\Pi_4$
vs. $\Pi_2$, with $R_{ap}/h^{-1}{\rm Mpc} = 0.4$. In a few cases the 
distributions of \M~ and data seem quite consistent or however strongly 
favoured, $e.g.$ for all plots at $R_{ap}/h^{-1}{\rm Mpc} = 0.8$. The 
probability of agreement, measured through KS coefficients, keeps low 
($\sim 10^{-1}$--10$^{-3}$), but the indication in favour of \M~ in respect to 
CDM is significant. We do not show analogous plots for \L, which turns out to 
be disfavoured in this test, based on gas distribution.

\begin{figure*}
\label{fig:track}
\centerline{\mbox{\epsfysize=16.0truecm\epsffile{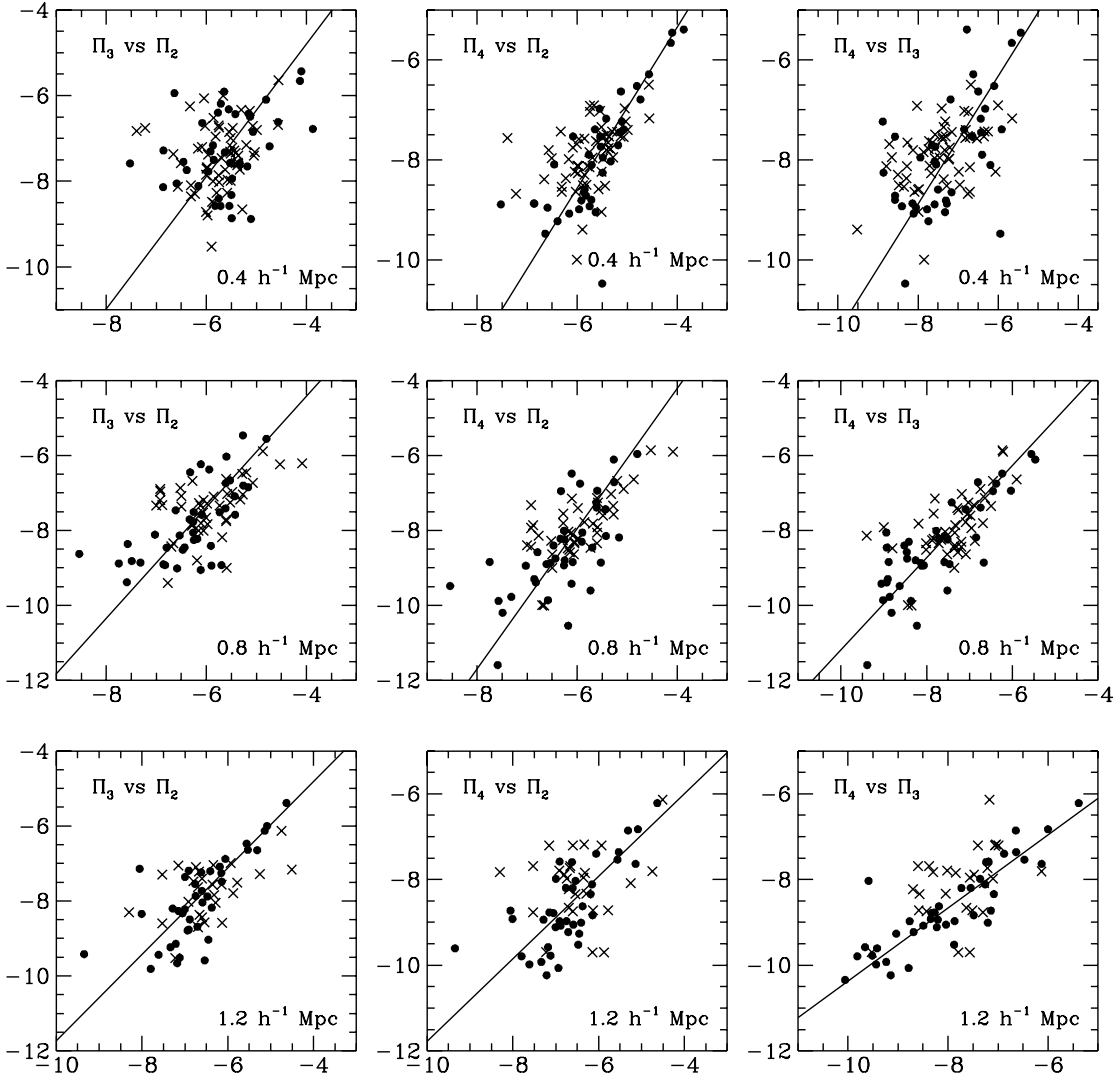}}}
\caption{Power ratios and evolutionary tracks for data (crosses)
and CHDM clusters (filled dots). The straight line is the best--fit 
to dots.}
\end{figure*}

\begin{figure*}
\label{fig:ang2}
\centerline{\mbox{\epsfysize=20.0truecm\epsffile{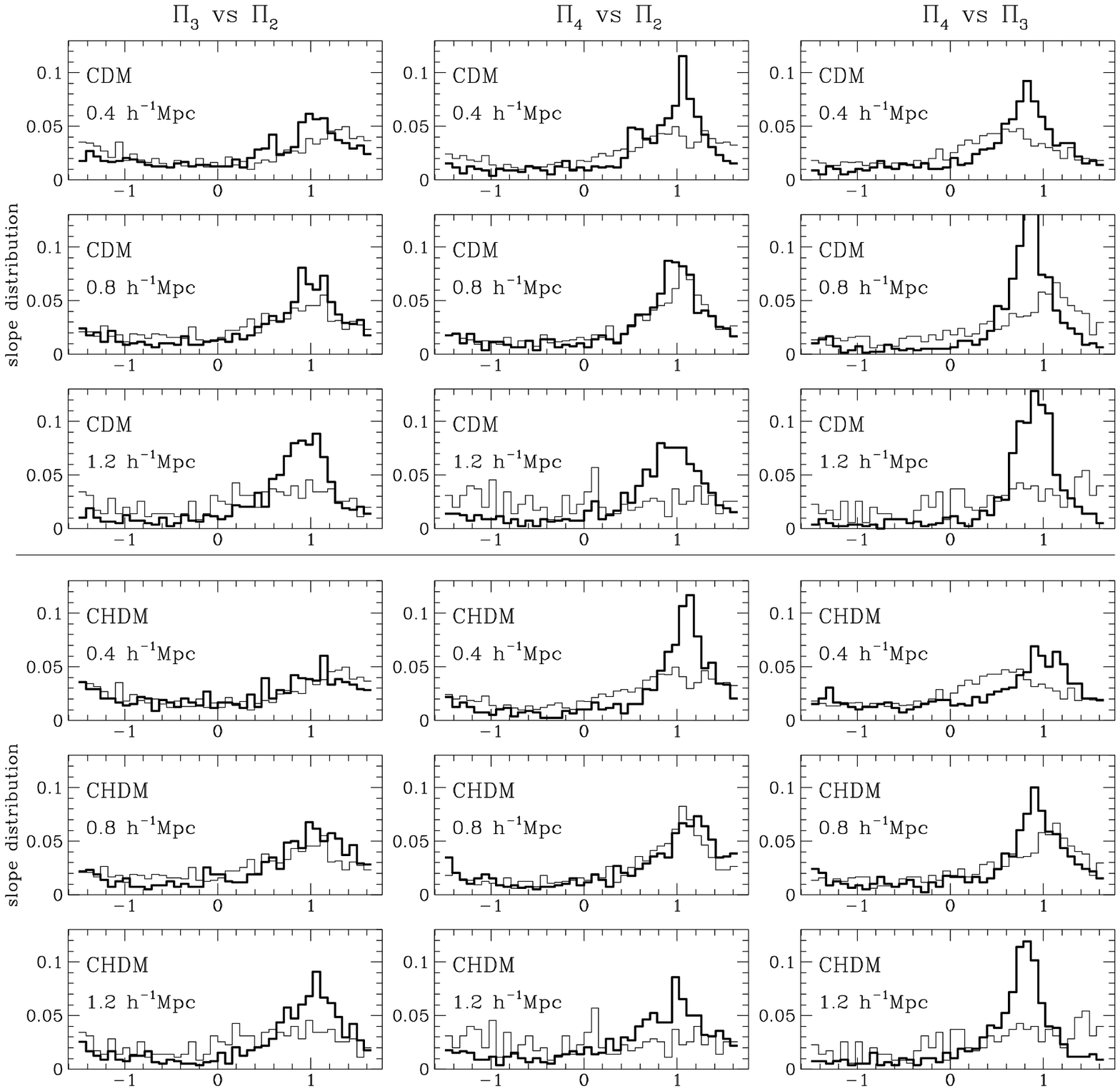}}}
\caption{Histogram in the range $(-\pi/2,\pi/2)$  of 
slope distributions. The thin line
is for data points; the thick line is for model (CDM or
\M) clusters.}
\end{figure*}

It would be unwise to draw any final conclusions on the basis
of this test. In our opinion, data need to be significantly improved and 
enriched before that this kind of analysis may become really discriminatory.
A possible pattern would amount to adding noise to model
data, seeking the level at which an agreement with real data is approached.
Hopefully, at this level, it may still be possible to discriminate among
different cosmologies. 

\subsection{Power ratios for gas and dark matter}
\label{sub:gasvsdark}

One of the results of this work concerns the reliability
of comparisons between data and simulations based on DM -- rather
than gas -- distributions. The quantitative outputs,
quite independently from the cosmological model, are that:
(i) DM \apim\ are systematically less correlated than gas \apim\
(DM slopes are more scattered than gas).
(ii) DM \apim\ are systematically greater than  gas \apim.

Both effects seem related to the increased complexity of the phase space 
distribution for a substance which is not constrained to be described by 
fluid variables. The first effect is however puzzling, in connection with 
the last point outlined in the previous subsection. As a matter of fact, 
DM correlation is closer to real data than gas correlation. In Fig.~9, 
for the sake of example, we compare the slope distributions for gas and DM.

In an attempt to find a reason for such
finding, we note that, at present,
galaxies may be expected to have a behaviour closer to DM particles 
rather than to gas. Although most of the $X$--ray emission is thought to 
come from baryons in the intracluster gas, elliptical galaxies could 
significantly contribute. As a matter of fact, at low redshifts, 
well resolved clusters show clearly a 
clumpy emission associated with galaxies. This should be taken into account, 
together with the fact that, in our SPH simulations, particle masses still 
exceed the average galaxy size, while gas cooling is not included.

These arguments, however, are only indicative. During their formation
stages, galaxies cannot be approximated by DM points and, to
account for their present distribution, we should consider a late DM input
arising from gas suppression, associated with cooling effects. In our
opinion, at this stage, the main conclusion is that this point is among
those to be explored through more resolved simulations, surely
including gas cooling and possibly other physical effects.

\begin{figure*}
\label{fig:angdmpbb}
\centerline{\mbox{\epsfysize=10.0truecm\epsffile{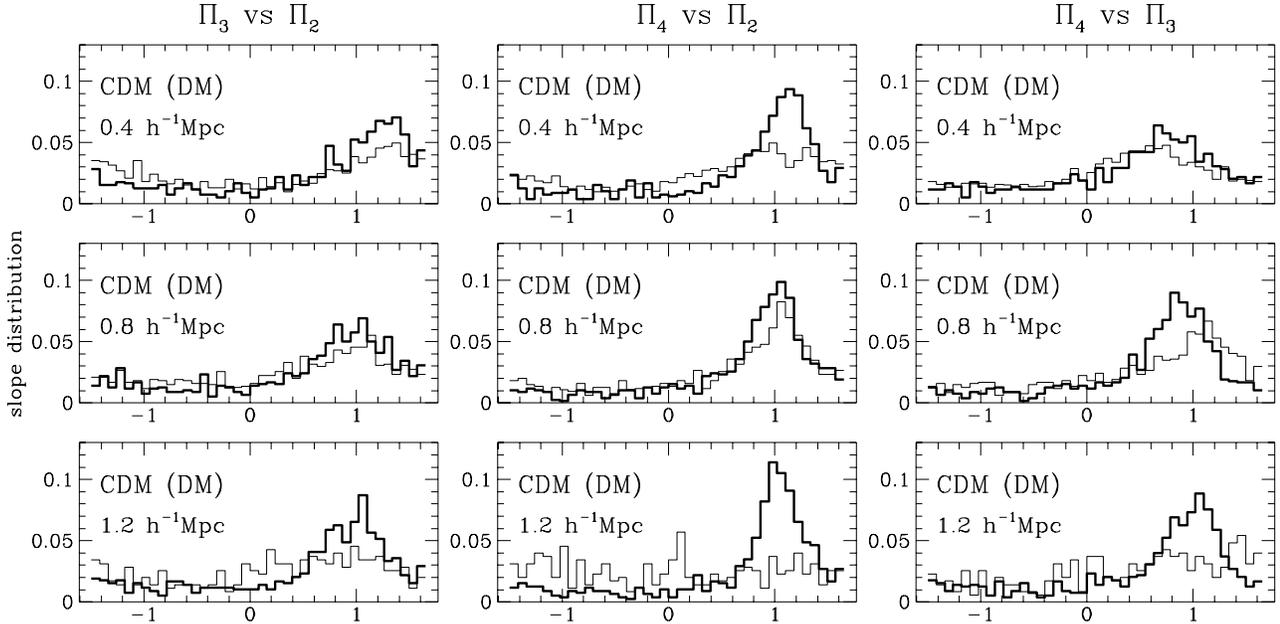}}}
\caption{Histogram in the range $(-\pi/2,\pi/2)$  of 
slope distribution for points worked out from gas distribution (thin line) 
and points worked out from dark matter distribution (thick line).}
\end{figure*}
However, if cosmological models are compared with data on the basis of
DM \apim, the conclusions drawn on the basis of Student t--test,
F--test and KS--test, are reversed. For the sake of example, $p$--$t$, 
$p$--$F$ and $p$--$KS$, for the comparison ROSAT--CDM and for $R_{ap} =
0.8 \, h^{-1}$Mpc, pass from 0.21 to 0.68$\cdot 10^{-3}$,
0.06 to 0.28 and 0.14 to 0.48$\cdot 10^{-2}$, for
$\Pi_2$; from  0.37 to 0.05, 0.11 to 0.97 and 0.18 to 0.22, for $\Pi_3$.
These shifts coherently indicate an increase in the
amount of substructures for DM with respect to the gas.

Altogether, DM \apim, unlike gas \apim, scarcely feel dissipative processes;
hence, using DM \apim, CDM and \M\ models
keep too many substructures and are no longer consistent with data;
on the contrary, the increase of substructures pushes \L\ to agree
with ROSAT sample outputs.

\section{Conclusions}
\label{sec:conclusions}

The global morphology of galaxy clusters is well described by
a multipole expansion of a pseudo--potential generating its
surface brightness, around its centroid. Such expansion leads to
the power ratios \apim, first introduced by BT95,
who also evaluated them for a ROSAT subsample.
In this work we have performed an extensive comparison of such
observational \apim, with \apim\ worked out from model clusters.
Such clusters were obtained for three cosmological models, all
normalized to present cluster number density. Such models are
CDM (which, therefore, is not consistent with COBE quadrupole data),
\L~ and \M~ with 20$\, \%$ of HDM given by a single neutrino flavour
(the latter two models are consistent with COBE quadrupole data).

Previous comparisons with simulated clusters were either restricted to 6
clusters, obtained with a TREESPH code by NFW, or used N--body non--hydro
simulations. Here, instead, we worked out 40 cluster models for each
cosmological model using a TREESPH code. 
We checked whether our hydro simulations have
a resolution adequate to yield safe \apim\ values
for the aperture radii considered. We tested that for a few
model clusters of the whole set, by changing softening and 
particle numbers. We found that \apim\ systematically show a greater
dependence on the choice of the line of sight, by which a model
cluster is assumed to be seen, than on variation of softening
or particle numbers. 

The hydro code does not take into account
gas cooling or supernova heating. This is expected to induce no bias, on the
scales we considered, 
provided that most of the observed
$X$--ray flux originates from the intracluster gas.

Among the results of our work, we wish to stress soon that \apim\ evaluated 
from gas distributions turn out to be substantially different from \apim\ 
worked out from DM distributions. This is to be ascribed to the smoothing 
effects of the interactions among gas particles, which erase anisotropies and 
structures. 
The effect is fairly significant and finally leads to a different score of
cosmological models in respect to data. In fact, while DM \apim\ worked out for
\L~ model clusters are close to data, this is no longer true for gas \apim\
within the same models. On the contrary, while DM \apim\ for CDM and \M~ models
show too much structure to fit data, gas \apim\ are in much better agreement
with data, for these models. Mohr \etal (1995) reached similar conclusions.
More precisely, they find that observed X--ray morphologies of clusters are
inconsistent with those obtained from a set of simulated clusters for
low--density CDM models. It is worth noticing that their simulations had been
performed using a combined N--body/hydro code (P3MSPH), while the BX set is
purely gravitational. It should be stressed that BX, after comparing ROSAT data
with N--body simulations, also outlined the need to extend the comparison to a
large set of hydrodynamical simulations. 

Here, the comparison between data and simulation was performed through
different steps. First of all, as BT96, TB and BX, we used the Student t--test,
the F--test and the KS--test to compare \apim\ distributions. We also 
considered the cluster distribution in the 3--dimensional parameter space 
with axes given by \apim\ ($m=2,3,4$), as well as projections of such 
distributions on planes.
Taking into account that data involve clusters at various redshifts, such
distribution provides an average evolutionary track for clusters. However,
comparing such distributions for data and models, we find a significantly
stronger correlation of \apim\ in models than in data. The degree of 
correlation depends on the model, but seems however in disagreement with data.
Model clusters tend to indicate a significantly faster evolution than data. The
cosmological model which seems closest to data is \M\ and it is possible that
different \M\ mixtures can lead to further improvements. It is also possible
that a slower evolution is an indication that the density of the Universe is
below critical. We therefore plan, in a near future, to perform cluster
simulations for \M~models with lighter neutrinos and for 0CDM models.

Let us recall that, at present, the main evidences in favour of 
open models
amount to the detections of large--scale matter concentrations at high 
redshift, either thanks to direct inspection 
(Bahcall $\&$ Fan 1998), or through the statistics of arcs arising from
lensing (Bartelmann et al. 1998).
The latter analysis, in particular, seems to exclude
a significant contribution of $\Lambda$ to the cosmic density,
in agreement with the findings of this work.
Evidences in favour of $\Lambda \neq 0$, instead, are mostly related
to recent improvements in the use of SNIa as
standard candles. They seem to support $\Omega = 1$ models with a large
$\Omega_\Lambda~(>$0.6, see Perlmutter et al. 1998 and
Reiss et al. 1998). 

Both such evidences seem however
to disagree with $\Omega$ estimates based on X--ray cluster surveys,
which may be consistent with $\Omega = 1$, without vacuum contribution
(see Sadat, Blanchard \& Oukbir 1998 and references therein)

In this work, the low probability of \L~models is also confirmed
by the analysis
of \apim\ correlations, which may become a discriminatory tool to test models. 
We have seen that separate distributions of \apim's can be in
fair agreement with data, while their joint distribution is not.

The disagreement found for such joint distributions also calls for an
improvement of the observational data set. In our work we used the same
observational set of previous analyses. We also tried to reproduce its possible
biases by reproducing the redshift distribution of data clusters in models
clusters. TB stressed that data selection is not based in cluster structure.
However, we have performed a further check to improve the safety of the use of
data. This amounted to test whether there is any correlation between cluster
luminosity and multipole structure. To do so, we shared our model clusters in 2
or 4 subsamples, ordered according to their intrinsical luminosity, and
verified that no correlation exists between \apim\ and luminosity. However,
before reaching final conclusions on the cosmological model through global
cluster features, an improvement of data is surely opportune.

Let us finally outline that the pseudo--potential, used to perform the 
multipole expansion, is somehow related to the potential yielding 
weak lensing; however, while lensing effects are related to the 
total density $\rho$, here we are referring to gas only, through its
square density $\rho_g^2$. The relation between $\rho$ and $\rho_g$
can be strongly model dependent, namely if the cosmic substance
contains fair amounts of late--derelativizing HDM, whose space
and velocity distribution, in non--linear structures, might be substantially
different from CDM itself. We therefore plan to do further work on model 
clusters to test relations between lensing effects and \apim\ expansion, 
also aiming to devise precise tests on DM composition.

\section*{Acknowledgments} 
One of us (SG) wishes to thank SISSA for its hospitality during 
the preparation of this work.
One of us (RV) wants to thank Y.P. Jing for an early critical
discussion on the aims of this work.
Ed Bertschinger, who refereed this paper, is to be
gratefully thanked for a number of important suggestions.

\end{document}